\documentclass[12pt]{article}
\usepackage{graphicx}

\newcommand{\eqn}{equation}
\newcommand{\dispst}{\displaystyle}
\newcommand{\scst}{\scriptstyle}
\newcommand{\scscst}{\scriptscriptstyle}
\newcommand{\cl}{\centerline}
\newcommand{\tworow}[2]{\stackrel{\scst #1}{\scst #2}}

\newcommand{\LRbk}[1]{\left( #1 \right)}
\newcommand{\LRbkM}[1]{\left\{ #1 \right\}}
\newcommand{\LRbkL}[1]{\left[ #1 \right]}
\newcommand{\LRbkN}[1]{\left. #1 \right.}
\newcommand{\LRcase}[1]{\left. #1 \right|}
\newcommand{\vs}{\,|\,}
\newcommand{\vect}[1]{\mbox{\boldmath$#1$}}
\newcommand{\va}[1]{\left| \vect{#1} \right|}
\newcommand{\vzero}{{\mathbf 0}}
\renewcommand{\Re}{{\rm Re}}
\renewcommand{\Im}{{\rm Im}}
\newcommand{\rd}{\partial}
\newcommand{\IntInf}[1]{\int_{-\infty}^{\infty}\!\! #1} 
\newcommand{\PIntInf}[1]{{\rm P}\!\!\!\int_{-\infty}^{\infty}\!\! {#1}} 
\newcommand{\PIntZI}[1]{{\rm P}\!\!\!\int_{0}^{\infty}\!\! {#1}} 
\newcommand{\fourier}[1]{\int\!\! d^{{\scscst #1}}\!x\, e^{ipx}}
\newcommand{\bra}[1]{\left\langle #1 \right|}
\newcommand{\ket}[1]{\left| #1 \right\rangle}
\newcommand{\abs}[1]{\left| #1 \right|}
\newcommand{\sgn}[1]{{\rm sgn}\!\left( #1 \right)}
\newcommand{\ave}[1]{\widehat{#1}}
\newcommand{\diff}[1]{\delta(#1)}
\newcommand{\ip}{\!\cdot\!}
\newcommand{\exchange}[2]{#1 \leftrightarrow #2}
\newcommand{\sqr}[1]{#1^{{\scscst 2}}}
\newcommand{\limzp}{\lim_{\epsilon \rightarrow +0}}
\newcommand{\order}{O}
\newcommand{\pp}{{p^{\scscst \prime}}}
\newcommand{\zero}[1]{{#1}^{\scscst 0}}
\newcommand{\pz}{\zero{p}}
\newcommand{\ppz}{\zero{\pp}}
\newcommand{\ieps}{i\epsilon}
\newcommand{\omg}{\omega}
\newcommand{\omgp}{{\omg^{\scscst \prime}}}
\newcommand{\omgsqr}{\omega^{2}}
\newcommand{\spr}{{s^{\scscst \prime}}}
\newcommand{\OpO}{{\mathcal O}}

\newcommand{\GeV}{\rm GeV}
\newcommand{\MeV}{\rm MeV}

\newcommand{\gm}{\gamma}
\newcommand{\gmf}{{\gamma^{5}}}
\newcommand{\sg}{\sigma}
\newcommand{\sgF}{\sigma^{5}}
\newcommand{\gmT}{\widetilde{\gamma}}
\newcommand{\sgT}{\widetilde{\sigma}}
\newcommand{\sgFT}{\widetilde{\sigma}^{5}}
\newcommand{\sla}[1]{{\scriptstyle \not} #1}
\newcommand{\lsla}[1]{\not\!\! #1}
\newcommand{\cd}{D}
\newcommand{\slacd}{\lsla{D}}
\newcommand{\lag}{{\mathcal L}}
\newcommand{\qb}{\overline{q}}
\newcommand{\muu}{m_{{u}}}
\newcommand{\md}{m_{{d}}}
\newcommand{\ms}{m_{{s}}}
\newcommand{\ma}{\widehat{m}}
\newcommand{\dmm}{\delta m}
\newcommand{\eu}{e_{{u}}}
\newcommand{\ed}{e_{{d}}}

\newcommand{\EV}[1]{\left\langle #1 \right\rangle}
\newcommand{\EVt}[3]{\left\langle #1 \left | #2 \right | #3 \right\rangle}
\newcommand{\EVs}[2]{\left\langle #1 \right\rangle_{\!#2}}
\newcommand{\EVV}[1]{\left\langle #1 \right\rangle_{\!\Vac}}
\newcommand{\EVM}[1]{\left\langle #1 \right\rangle_{\!\Med}}
\newcommand{\EVq}[1]{\left\langle q \!\left| #1 \right|\! q \right\rangle}

\newcommand{\qc}[1]{\overline{#1} #1}
\newcommand{\qcv}[2]{\overline{#1} #2 #1}
\newcommand{\qn}[1]{{#1}^{\dagger} #1}
\newcommand{\qnv}[2]{{#1}^{\dagger} #2 #1}
\newcommand{\Parity}{{\mathcal P}}
\newcommand{\TimeRev}{{\mathcal T}}
\newcommand{\pT}{\widetilde{p}}
\newcommand{\qT}{\widetilde{q}}
\newcommand{\xT}{\widetilde{x}}
\newcommand{\Corr}[3]{\Pi^{#1}_{#2} \left( #3 \right)}
\newcommand{\Corrp}[2]{\Pi^{#1}_{#2}}
\newcommand{\CorrE}[3]{\Pi^{#1 {\rm (E)}}_{#2} \left( #3 \right)}
\newcommand{\CorrO}[3]{\Pi^{#1 {\rm (O)}}_{#2} \left( #3 \right)}
\newcommand{\CorrEp}[2]{\Pi^{#1 {\rm (E)}}_{#2}}
\newcommand{\CorrOp}[2]{\Pi^{#1 {\rm (O)}}_{#2}}
\newcommand{\Spec}[3]{\rho^{#1}_{#2} \left( #3 \right)}

\newcommand{\SpecT}[3]{{\widetilde{\rho}}^{#1}_{#2} \left( #3 \right)}

\newcommand{\SpecG}[3]{\Delta^{#1}_{#2}\!\left( #3 \right)}
\newcommand{\SpecGp}[2]{\Delta^{#1}_{#2}}
\newcommand{\Vo}{V_{1}}
\newcommand{\Vt}{V_{2}}
\newcommand{\Ao}{A_{1}}
\newcommand{\At}{A_{2}}
\newcommand{\To}{T_{1}}
\newcommand{\Tt}{T_{2}}

\newcommand{\csop}[1]{\lambda_{#1}}
\newcommand{\MMEM}{\Theta}
\newcommand{\MMEMb}{\overline{\Theta}}
\newcommand{\MME}[1]{\theta^{#1}}
\newcommand{\M}{M}
\newcommand{\Ma}{\widehat{\M}}
\newcommand{\DM}{{\mit \Delta} \!\hspace{0.1em} \M}

\newcommand{\Lint}{\lag_{\rm int}}
\newcommand{\Hint}{H_{\rm int}}
\newcommand{\HQCD}{{\mathcal{H}}^{\QCD}}
\newcommand{\MaN}{\widehat{\M}_{N}}
\newcommand{\dMN}{\delta \!\hspace{0.06em} \M_{N}}
\newcommand{\bary}[1]{\Psi_{#1}}
\newcommand{\baryb}[1]{\anti{\Psi}_{#1}}

\newcommand{\AngleP}{\theta}
\newcommand{\AngleA}{\overline{\theta}}

\newcommand{\CoeffSL}{\frac{2}{\sqrt{3}}}
\newcommand{\currn}[1]{\eta_{#1}}
\newcommand{\currbn}[1]{\overline{\eta}_{#1}}
\newcommand{\curr}[2]{\eta_{#1}\!\left(#2\right)}
\newcommand{\currb}[2]{\overline{\eta}_{#1}\!\left(#2\right)}
\newcommand{\smz}{m_{{0}}^{2}}

\newcommand{\Thr}{S_{\scscst 0}}
\newcommand{\ThrE}[1]{\Thr^{#1 ({\rm E})}}
\newcommand{\ThrO}[1]{\Thr^{#1 ({\rm O})}}
\newcommand{\BetaSL}{\beta_{\Lambda\Sigmaz}}
\newcommand{\Borel}{{\mathcal B}}
\newcommand{\Msq}{M^{2}}
\newcommand{\Efunc}[2]{E_{#1}\!\LRbk{#2}}
\newcommand{\ffunc}[2]{f^{#1}\!\LRbk{#2}}
\newcommand{\Ffunc}[2]{F^{#1}\!\LRbk{#2}}
\newcommand{\Ftfunc}[2]{\widetilde{F}^{#1}\!\LRbk{#2}}
\newcommand{\MSz}{\M_{\Sigmaz}}
\newcommand{\ML}{\M_{\Lambda}}
\newcommand{\MSzsq}{\sqr{\M}_{\Sigmaz}}
\newcommand{\MLsq}{\sqr{\M}_{\Lambda}}
\newcommand{\Masq}{\sqr{\Ma}}
\newcommand{\Fn}{{\rm I}}
\newcommand{\FP}{{\rm Ia}}
\newcommand{\FC}{{\rm Ib}}
\newcommand{\Fd}{{\rm II}}
\newcommand{\FdP}{{\rm IIa}}
\newcommand{\FdC}{{\rm IIb}}
\newcommand{\EM}{{\rm EM}}
\newcommand{\QCD}{{\rm QCD}}
\newcommand{\OPE}{{\rm O\!\hspace{0.06em}P\!\hspace{0.06em}E}}
\newcommand{\Phen}{{\rm P\!\hspace{0.06em}h\!\hspace{0.09em}e\hspace{0.0em}n}}
\newcommand{\Pole}{{(}{\rm pole}{)}}
\newcommand{\Cont}{{(}{\rm cont}{)}}
\newcommand{\Phys}{{\rm P\!\hspace{0.06em}hys}}

\newcommand{\Vac}{{\scscst 0}}
\newcommand{\Med}{{\scscst\rm M\!\hspace{0.05em}e\!\hspace{0.09em}d}}
\newcommand{\Dens}[1]{\rho_{#1}}
\newcommand{\DensT}{\rho_{\!\hspace{0.06em}\scscst N}}
\newcommand{\dDens}{\delta \rho_{\!\hspace{0.06em}\scscst N}}
\newcommand{\DensS}{\rho_{\scscst 0}}
\newcommand{\Asymm}{\alpha_{np}}

\newcommand{\anti}[1]{\overline{#1}}
\newcommand{\Sigmaz}{\Sigma^{\scscst 0}}
\newcommand{\rhoz}{\rho^{\scscst 0}}
\newcommand{\pion}[1]{\pi^{\scscst #1}}
\newcommand{\etapr}{\eta^{\scscst \prime}}
\newcommand{\szl}{\Sigmaz-\Lambda}
\newcommand{\antiszl}{\anti{\Sigmaz}-\anti{\Lambda}}
\newcommand{\rzo}{\rhoz-\omega}

\newcommand{\FigPhendia}[2]{
    \begin{figure}[#1]
        \begin{center}
            \includegraphics*[#2]{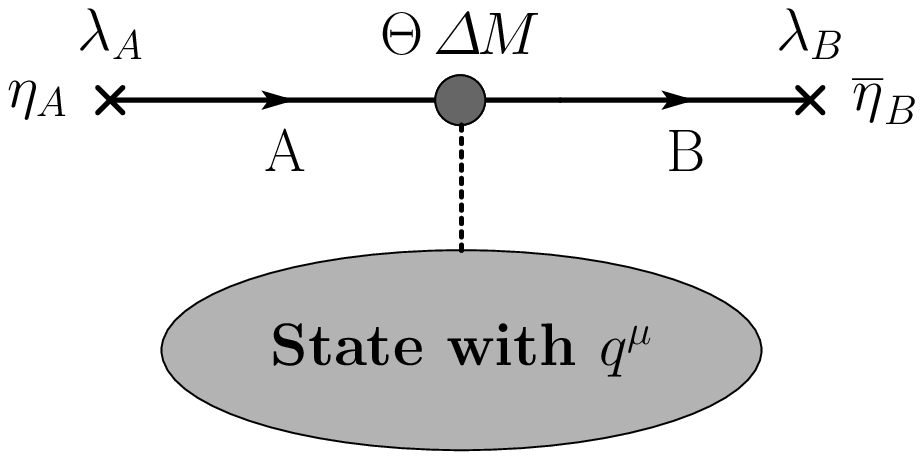}
        \end{center}
        \caption{
            A diagrammatic illustration of Eq.(\ref{eqn:CorrTPole}).
            The baryon $A\,(B)$ couples to 
            the interpolating operator $\currn{A}\,(\currbn{B})$ 
            with the coupling strength $\csop{A}\,(\csop{B})$.
            The baryon mixing  $\MMEM \!\,\, \DM$ 
            is induced by 
            the state with four-momentum $q^{\mu}$ 
            which will be later identified 
            with the isospin-asymmetric nuclear medium.
        }
        \label{fig:Phendia}
    \end{figure}
}
\newcommand{\FigOPEdia}[2]{
    \begin{figure}[#1]
        \begin{center}
            \includegraphics*[#2]{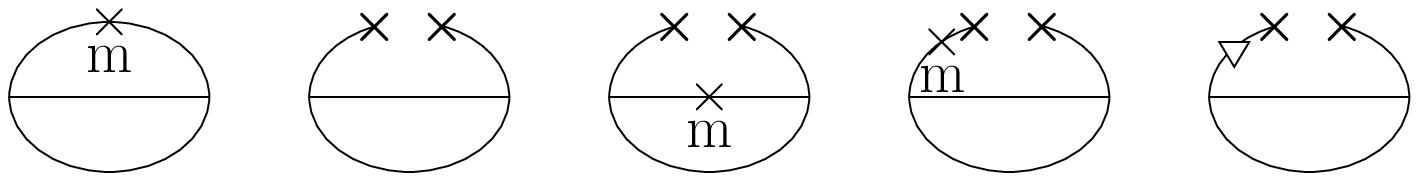}
        \end{center}
        \caption{
            Diagrams of OPE up to dimension 4 for the $\szl$ mixied 
            correlation function.
            (m stands for the $u,d,s$ quark masses and 
            $\bigtriangledown$ for the covariant derivative.)
        }
        \label{fig:OPEdia}
    \end{figure}
}
\newcommand{\FigBSRmAngleS}[1]{
    \begin{figure}[#1]
        \begin{center}
            \includegraphics*[width=0.5\linewidth]{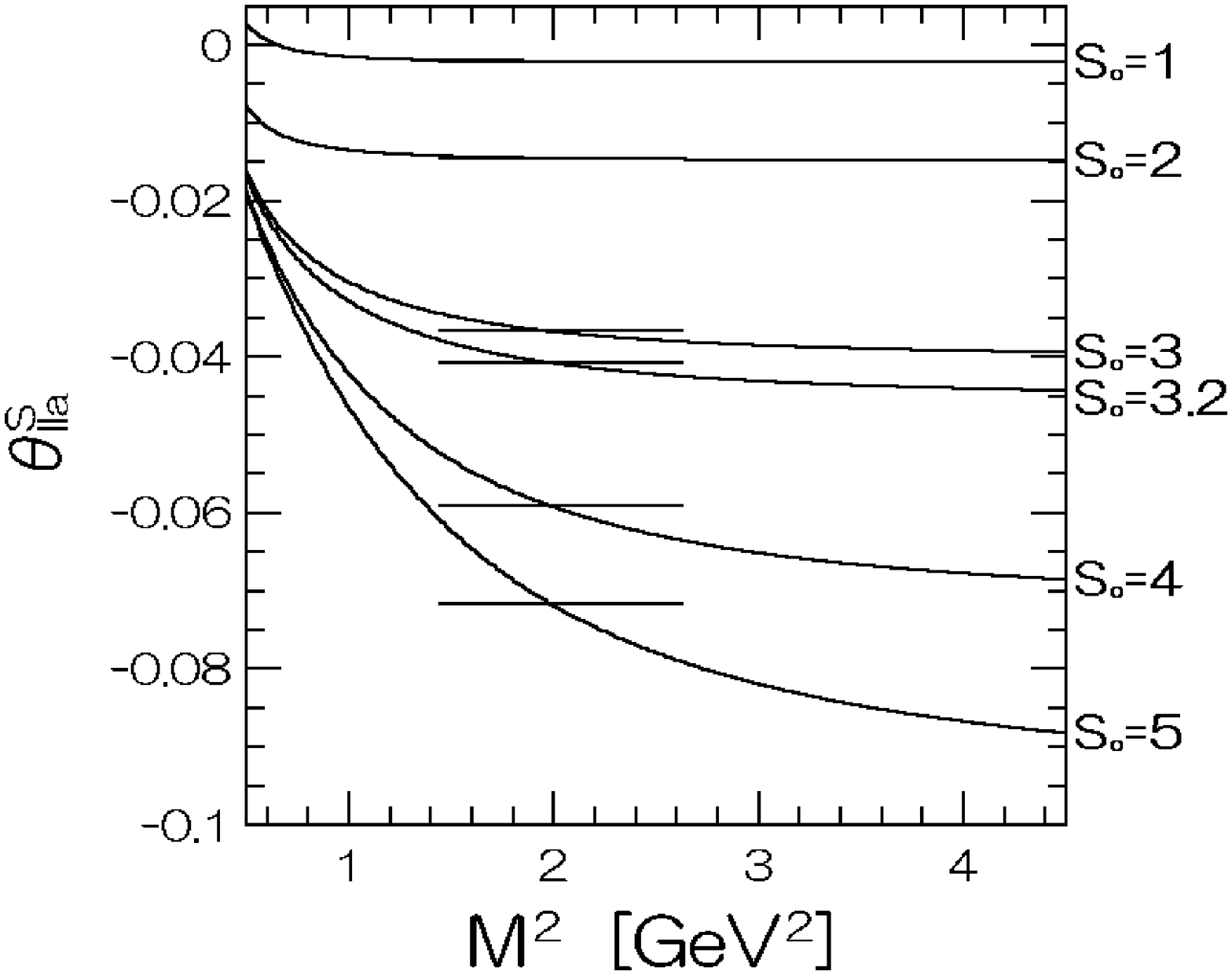}
        \end{center}
        \vspace{-5mm}
        \caption{
            The scalar angle in the Type {\FdP} sum rule is shown
            as a function of the Borel mass $M^2$ 
            for different values of the threshold $\Thr\,[\GeV^{2}]$.
            The straight horizontal lines imply 
            the Borel window and the averaged value in the Window.
            $\BetaSL=2.5\,{\GeV}^{6}$ and $ \dmm=3.9{\MeV}$ are used.
        }
        \label{fig:BSRmAngleS}
    \end{figure}
}
\newcommand{\FigBSRAngleS}[1]{
    \begin{figure}[#1]
        \begin{center}
        \ \ \ 
        \includegraphics*[width=0.94\linewidth]{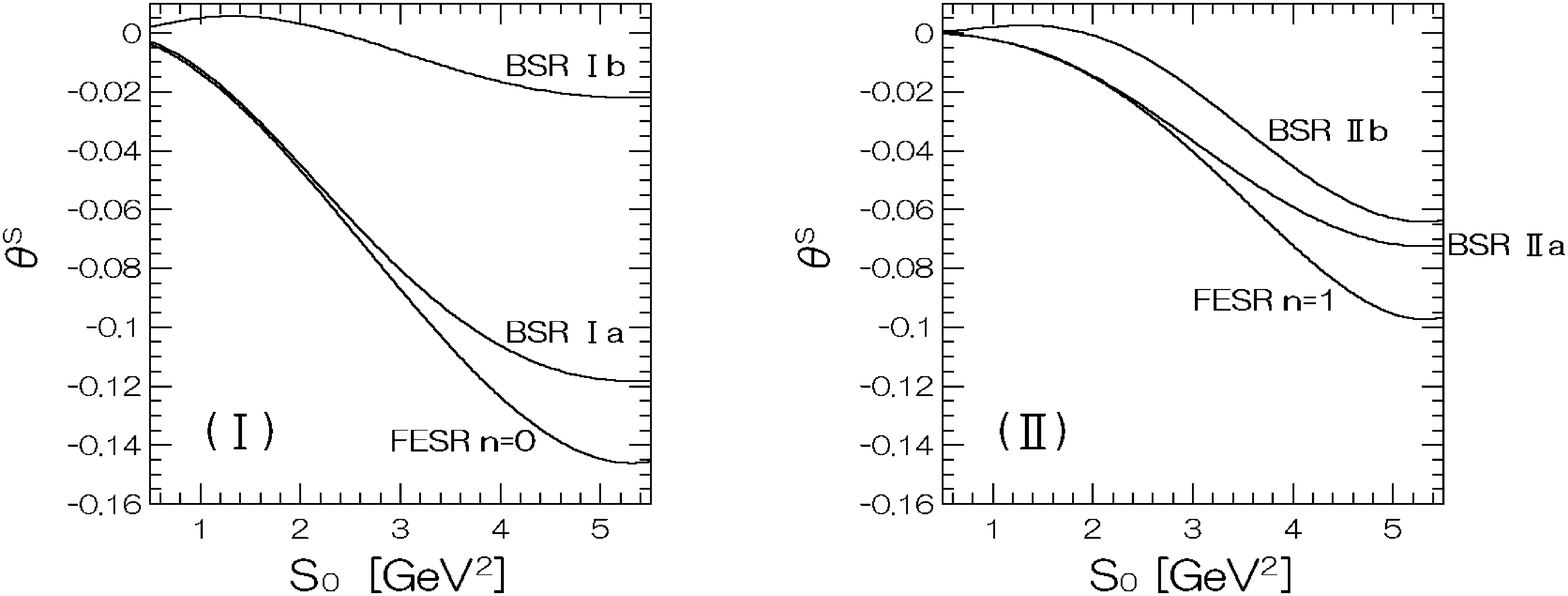}
        \end{center}
        \vspace{-5mm}
        \caption{
            The scalar angle is shown as a function of 
            the continuum threshold $\Thr\,[\GeV^{2}]$.
            The left panel is for BSR Type {\Fn} 
            and the right panel is for BSR Type {\Fd} 
            with corresponding $n$-th order FESR.
            $\BetaSL=2.5\,{\GeV}^{6}$ and $\dmm=3.9{\MeV}$ are used.
        }
        \label{fig:BSRAngleS}
    \end{figure}
}
\newcommand{\FigBSRAngleV}[1]{
    \begin{figure}[#1]
        \begin{center}
            \ \ \ 
            \includegraphics*[width=0.94\linewidth]{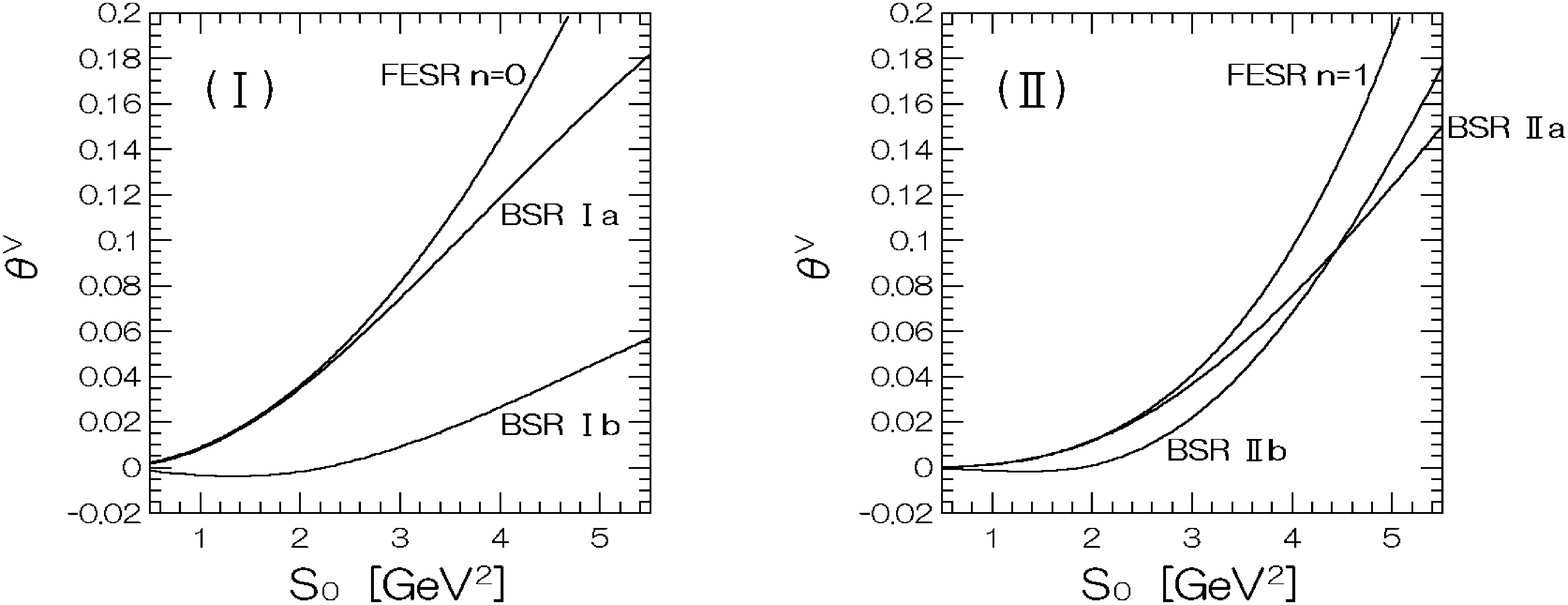}
        \end{center}
        \vspace{-5mm}
        \caption{
            The vector angle is shown as a function of the continuum
            threshold $\Thr\,[\GeV^{2}]$.
            The left panel is for BSR Type {\Fn} 
            and the right panel is for BSR Type {\Fd} 
            with corresponding $n$-th order FESR.
            $\BetaSL=2.5\,{\GeV}^{6}$ and $\dmm=3.9{\MeV}$ are used.
        }
        \label{fig:BSRAngleV}
    \end{figure}
}
\newcommand{\FigBSRAngleFdP}[1]{
    \begin{figure}[#1]
        \begin{center}
            \ \ \ 
            \includegraphics*[width=0.94\linewidth]{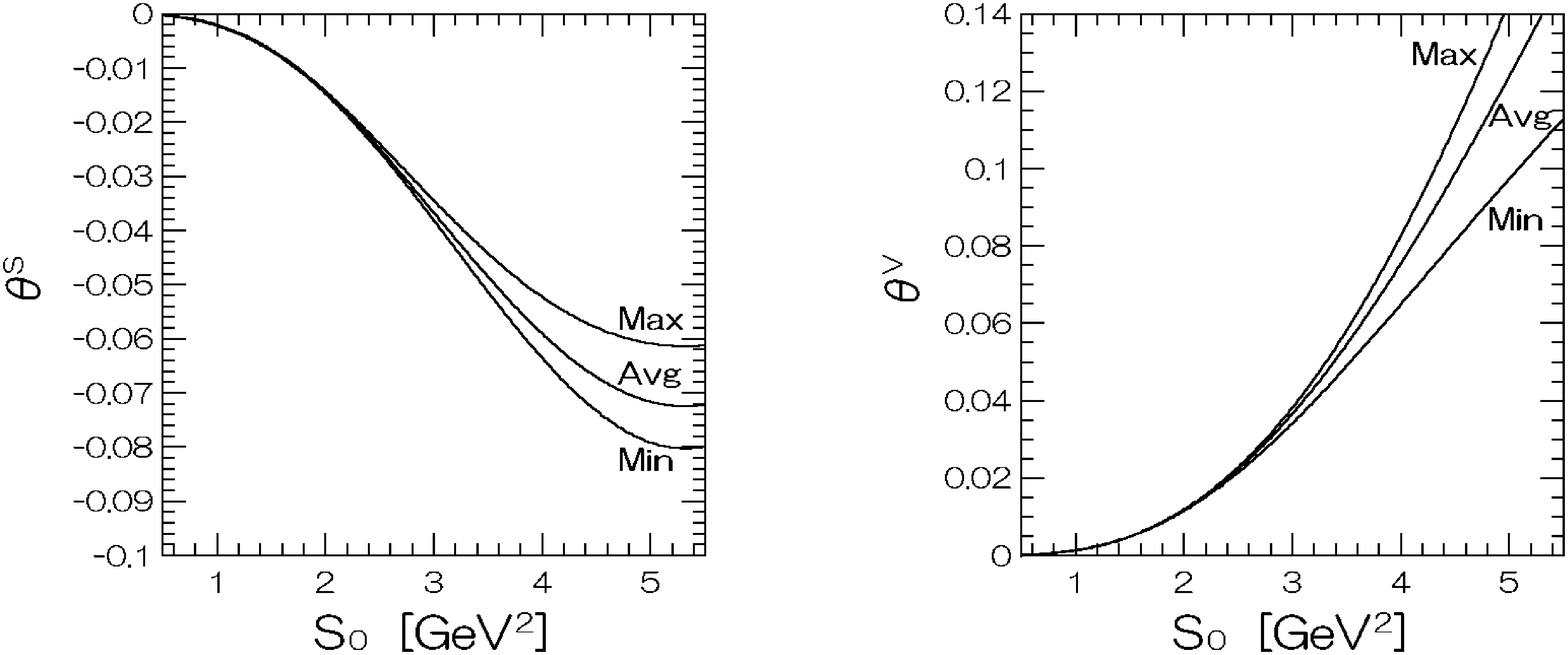}
        \end{center}
        \vspace{-5mm}
        \caption{
            The scalar and vector angles are shown as a function of 
            the continuum threshold $\Thr\,[\GeV^{2}]$ for Type {\FdP}.
            The curve with the label ``Avg" is 
            the average of the sum rule over the Borel window.
            The curve ``Max"\,(``Min") is 
            the maximum\,(minimum) value of the sum rule 
            in the Borel window.
        }
        \label{fig:BSRAngleFdP}
    \end{figure}
}
\newcommand{\FigBSRAsymmFdP}[1]{
    \begin{figure}[#1]
        \begin{center}
            \ \ \ 
            \includegraphics*[width=0.94\linewidth]{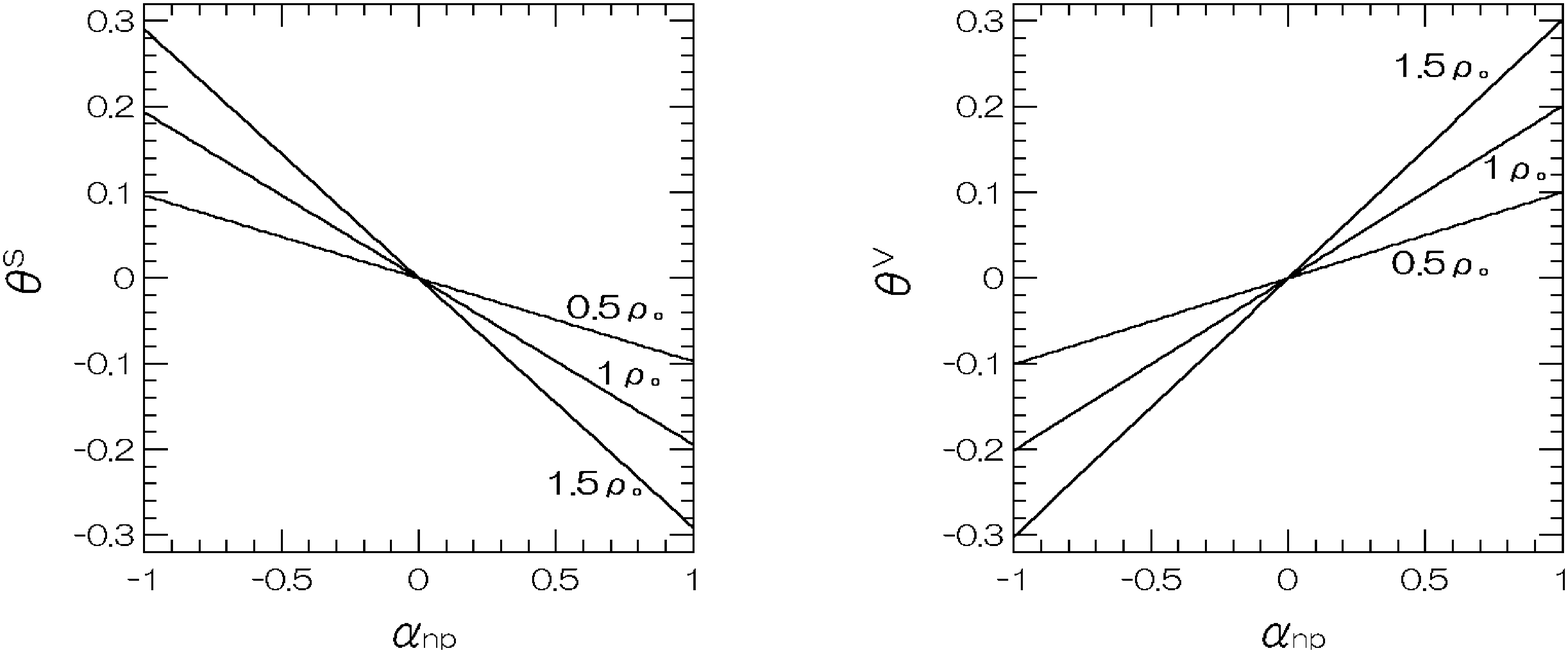}
        \end{center}
        \vspace{-5mm}
        \caption{
            The mixing angles (Type \FdP) are shown as a function of 
            the $n-p$ asymmetry $\Asymm\,(=\dDens/\DensT)$ 
            for different values of the total density $\DensT$.
        }
        \label{fig:BSRAsymmFdP}
    \end{figure}
}
\newcommand{\TbQCDparameter}[1]{
    \begin{table}[#1]
        \begin{center}
            \begin{tabular}{||c|c||}
                \hline 
                $ \ma\,\, (=(\muu +\md)/2) $ 
                & $ 7.7\,{\MeV} $ \\
                \hline
                $ \dmm /\ma\,\, (\dmm =\md-\muu) $ 
                & $ 0.51 $ \\
                \hline
                $ \ms /\ma $
                & $ 19 $ \\
                \hline \hline
                $ \EVV{\ave{\qc{q}}}\,\, (=\EVV{\qc{u}+\qc{d}}/2) $ 
                & $ (-275\,{\MeV})^{3} $ \\
                \hline
                $ \gamma\,\, (=\EVV{\qc{d}}/\EVV{\qc{u}}-1) $ 
                & $ -5.6 \times 10^{-3} $ \\
                \hline
                $ \beta\,\, (=\EVV{\qc{s}}/\EVV{\qc{u}}-1) $ 
                & $ -0.22 $ \\ 
                \hline \hline
                $ \EVV{\frac{\alpha_{\!s}}{\pi} G^{2}} $ 
                & $ (347\,{\MeV})^{4} $ \\
                \hline
                $ \smz\,\,(=-g_{\!s}\!\EVV{\qcv{q}{(\sg \ip G)}}
                /\EVV{\qc{q}})$ 
                & $ 0.91\,{\GeV}^{2} $ \\
                \hline
            \end{tabular}
        \end{center}
        \caption{
            QCD parameters in the vacuum 
            at the renormalization scale $1\,{\GeV}^{2}$. 
            Those are determined to reproduce 
            the octet baryon spectrum \cite{NYM}.
        }
    \label{tb:QCDparameter}
    \end{table}
}
\begin{document}
\setcounter{equation}{0}
\setcounter{table}{0}
\setcounter{figure}{0}

\vspace{1.5cm}
\vspace{5pt}
\cl{
    {\Large \bf In-medium $\szl$ Mixing in QCD Sum Rules}
}
\vspace{1.0cm}
\cl{
    N. Yagisawa\footnote{e-mail address: yagisawa@nucl.ph.tsukuba.ac.jp, 
        yagisawa@nt.phys.s.u-tokyo.ac.jp}
}
\vspace{0.3cm}
\cl{
    {\em Institute of Physics, University of Tsukuba, Tsukuba, 
    Ibaraki 305-8571, Japan}
}
\vspace{0.5cm}
\cl{
    T. Hatsuda\footnote{e-mail address: hatsuda@phys.s.u-tokyo.ac.jp}
    { and }
    A. Hayashigaki\footnote{e-mail address: arata@nt.phys.s.u-tokyo.ac.jp}
}
\vspace{0.3cm}
\cl{
    {\em Physics Department, University of Tokyo, Tokyo 113-0033, Japan}
}
\vspace{1.0cm}
\cl{\today}
\vspace{2.0cm}
\begin{abstract}
    The $\szl$ mixing angle 
    in isospin-asymmetric nuclear medium 
    is investigated by using QCD sum rules.
    From the general consideration of the 
    in-medium baryonic correlations,
    in-medium baryon mixings are shown to have 
    several Lorentz structures 
    such as the scalar mixing angle $\MME{S}$ and 
    the vector mixing angle $\MME{V}$.
    This causes a difference between 
    the particle mixing $\AngleP \,(= \MME{S} + \MME{V})$ 
    and the  anti-particle mixing $\AngleA \,(= \MME{S} - \MME{V})$.
    From the finite energy sum rules for the $\szl$ mixing,
    we find that the in-medium part of the mixing angle 
    has a relation $\MME{S}_{\Med} \sim - \MME{V}_{\Med}$
    in the isospin-asymmetric medium. 
    This implies that the medium affects mainly the anti-particle mixing.
    From the Borel sum rules, we obtain 
    $|\,\AngleA - \AngleP_{\Vac}| \simeq 
    0.39 \,|(\Dens{n} - \Dens{p})|/\DensS$
    with $\AngleP_{\Vac}$, $\Dens{n}$, $\Dens{p}$ and $\DensS$ 
    being the vacuum mixing angle, 
    the neutron density, the proton density and 
    the normal nuclear matter density respectively. 
\end{abstract}

\vspace{1.0cm}
{\footnotesize PACS number(s): 11.55.Fv, 11.55.Hx, 12.38.-t, 
    14.20.-c, 14.20.Jn, 21.65.+f}

%
%

\setcounter{footnote}{0}

\newpage
\section{INTRODUCTION}
\setcounter{equation}{0}
\label{sec:Intro}

The $SU(2)$ isospin symmetry is slightly broken in the hadronic world.
Examples of this symmetry breaking are 
the mass splittings within a same isospin multiplet 
($p-n$, $\Sigma^{\pm}-\Sigmaz$ and $\pion{\pm}-\pion{0}$), 
the particle mixing among different isospin multiplets 
($\pion{0}-\eta$, $\pion{0}-\etapr$, $\rzo$ and $\szl$), 
and the nuclear force in the ${}^{1\!}S_{0}$ channel 
($V_{pp} \neq V_{pn} \neq V_{nn}$) \cite{MNS,H/OPTW}.

The isospin symmetry breaking has two different sources: 
(i) the electromagnetic (EM) effect 
due to the electric-charge difference between 
$u$ and $d$ ($\eu \neq \ed$), 
and (ii) the quark-mass difference between 
$u$ and $d$ ($\muu \neq \md$).
The latter effect can be evaluated from 
the mass term $\HQCD_{\rm mass}$ of the QCD Hamiltonian density 
for light flavors, 
\begin{\eqn}
    \HQCD_{\rm mass} = \frac{1}{2}(\muu+\md)(\qc{u}+\qc{d}) 
        + \frac{1}{2}(\md-\muu)(\qc{d}-\qc{u}) 
        + m_s \bar{s}s ,
\end{\eqn}
where $(\md-\muu)/(\muu+\md) \sim 0.29\ \cite{GL}$. 
$\HQCD_{\rm mass}$ is known to be 
more important than the EM effect 
for the $p-n$ mass difference, the $\rzo$ mixing, 
and the $\szl$ mixing \cite{GTW}.

The QCD sum rule \cite{SVZ} is a useful method 
to evaluate the magnitude of the isospin symmetry breaking 
with non-perturbative QCD dynamics.
It has been applied for 
the isospin mass splittings in octet baryons \cite{HHP,YHHK,ADI}, 
the $\rzo$ mixing \cite{SVZ,HHMK}, 
the $\pion{0}-\eta$ mixing \cite{CHM} and 
the $\szl$ mixing \cite{ZHY,NYM}.
For example, the $\szl$ mixing angle in the vacuum defined by 
\begin{\eqn}
\AngleP_{\Vac} = -\frac{\EVt{\Sigmaz}{H^{\QCD}_{\rm mass}}{\Lambda}}
{\M_{\Sigmaz}-\M_{\Lambda}}
\end{\eqn}
is evaluated as $\abs{\AngleP_{\Vac}} = 1.4 \times 10^{-3}$ \cite{NYM} and
$7 \times 10^{-3}$ \cite{ZHY} in QCD sum rules.
This value is comparable to the other estimate 
$|\AngleP_{\Vac} | \simeq 1 \times 10^{-2}$ 
in the naive quark model \cite{ISGUR} 
and in the chiral perturbation theory \cite{GL,GTW}.

In this paper, we will consider 
the isospin-asymmetric nuclear medium where 
the difference between 
the neutron density ($\Dens{n}$) and the proton density ($\Dens{p}$) 
becomes an extra source of the isospin symmetry breaking.
In particular, we study how this new source affects the $\szl$ mixing.
The in-medium QCD sum rule \cite{HL,KH,CFGJ} 
is a suitable method for this purpose, 
since we can treat the isospin-asymmetric medium 
as a background field acting on the $\szl$ correlation 
through the operator product expansion.
Also, it allows us to investigate 
the response of the mixing angle 
under the variation of the magnitude of isospin-asymmetry.

We should mention here that
the $\rzo$ mixing in isospin-asymmetric medium 
has been recently studied in \cite{DHP}.
A major difference between the meson-mixing treated in \cite{DHP} and 
the baryon-mixing in the present paper lies 
in the fact that the latter can have several Lorentz structures 
(such as scalar and vector mixing angles) 
in the medium because of the spinor structure of the baryon fields.
This will cause an interesting difference 
between the  particle mixing ($\szl$) 
and the anti-particle mixing ($\antiszl$).
In the former (latter), the scalar mixing and the 
vector mixing act in destructive (constructive) manner. 
A close analogy of this phenomenon is 
the scalar and vector self-energies of the nucleon (anti-nucleon) 
in the symmetric nuclear medium, 
where scalar and vector act in destructive (constructive) way \cite{SW}.

The organization of this paper is as follows.
In Sec.\ref{sec:GPCorr}, 
we analyze the general structure of in-medium correlation functions 
of spin$\,\frac{1}{2}$ baryonic currents with and without the mixing.
Dispersion relations satisfied by the correlation functions 
are also written down after decomposing them into 
even and odd parts with respects to the frequency $\omega$ of the currents.
In Sec.\ref{sec:MixForm}, 
we introduce a generalized mixing matrix in the spinor space 
and make physical interpretation of 
the scalar and vector mixing angles.
In Sec.\ref{sec:OPEinMed}, 
we carry out the operator product expansion (OPE) 
of the mixed correlation function in the $\szl$ channel.
The Lorentz-tensor and isospin-asymmetric operators are kept in OPE 
since they have non-vanishing expectation values 
in the isospin-asymmetric medium.
The in-medium condensates which appear in OPE 
are also evaluated in the low density expansion in this section.
Since the in-medium expectation values of 
the isospin-asymmetric operators beyond dimension 4 
are hard to be determined at present, 
we limit ourselves to the OPE up to dimension 4.
In Sec.\ref{sec:QSR}, 
we construct the finite energy sum rules \cite{KPT} 
and the Borel sum rules \cite{SVZ} 
using the results in previous sections.
Then we extract a qualitative result from 
the finite energy sum rule.
In Sec.\ref{sec:NumA}, 
to reduce the uncertainties due to 
the absence of higher dimensional operators in OPE, 
we examine the reliability of the sum rules 
constructed in Sec.\ref{sec:QSR} from the point of view of 
the consistency among different sum rules.
Then we evaluate the $\szl$ mixing angles numerically.
Sec.\ref{sec:Conclud} is devoted to 
summary and concluding remarks.

\setcounter{equation}{0}
\section{GENERAL PROPERTIES OF THE CORRELATION FUNCTIONS}
\label{sec:GPCorr}

In this section we examine the spinor structures of 
the diagonal and off-diagonal correlation functions of 
spin$\,\frac{1}{2}$ baryonic currents.
We will also derive the dispersion relation for each spinor component 
of the correlation functions.

\subsection{Spinor structure}
\label{sec:GPCorrSS}

Let us start with the following two-point functions: 
\begin{eqnarray}
    \Corr{}{T}{p \vs q} & = & i\fourier{4}\EVq{T[\curr{A}{x}\currb{B}{0}]}, 
    \label{eqn:corrabMT}
    \\
    \Corr{}{R}{p \vs q} & = & i\fourier{4}\EVq{R[\curr{A}{x}\currb{B}{0}]}, 
    \label{eqn:corrabMR} 
\end{eqnarray}
where $T$ and $R$ denote time-ordered and retarded products respectively, 
and $\curr{A(B)}{x}$ is an interpolating operator for the baryon $A\,(B)$. 
If $A$ is different from $B$, the correlations describe the particle mixings.
$\ket{q}$ is a state vector with four-momentum $q^{\mu}$.
Later, this state will be identified with 
the isospin-asymmetric nuclear medium 
to investigate the $\szl$ mixing with $A=\Lambda$ and $B=\Sigmaz$. 

These correlation functions have 
the following spectral representations, 
\begin{eqnarray}
    \Corr{}{T\,\alpha\beta}{p \vs q} & = & \frac{1}{2\pi} \IntInf{d\ppz} 
        \LRbkL{\frac{\Spec{}{\alpha\beta}{\pp \vs q}}{\ppz-\pz-\ieps}
            + \frac{\SpecT{}{\alpha\beta}{\pp \vs q}}{\ppz-\pz+\ieps}},
    \\
    \Corr{}{R\,\alpha\beta}{p \vs q} & = & \frac{1}{2\pi} \IntInf{d\ppz} 
        \LRbkN{\frac{\Spec{}{\alpha\beta}{\pp \vs q}
            + \SpecT{}{\alpha\beta}{\pp \vs q}}{\ppz-\pz-\ieps}},
\end{eqnarray}
with $p^{\mu}=(\pz,\vect{p})$ and $\pp^{\mu}=(\ppz,\vect{p})$.
$\Spec{}{\alpha\beta}{p \vs q}$ and $ \SpecT{}{\alpha\beta}{p \vs q}$ are 
the spectral functions defined by
\begin{eqnarray}
    \Spec{}{\alpha\beta}{p \vs q} & = & \fourier{4}
        \EVq{\LRbkM{\curr{A}{x}}_{\alpha}\LRbkM{\currb{B}{0}}_{\beta}}, 
    \label{eqn:Specf}
    \\
    \SpecT{}{\alpha\beta}{p \vs q} & = & \fourier{4}
        \EVq{\LRbkM{\currb{B}{0}}_{\beta}\LRbkM{\curr{A}{x}}_{\alpha}},
    \label{eqn:SpecfT}
\end{eqnarray}
where spinor indices ($\alpha,\beta$) are explicitly written.

To make the following discussion concise, 
let us introduce a linear combination of the spectral functions 
with real parameters $a$ and $b$ as 
\begin{\eqn}
    \SpecG{}{}{p \vs q} \equiv 
        a \, \Spec{}{}{p \vs q} + b \, \SpecT{}{}{p \vs q}
    \label{eqn:DefSpecG}
\end{\eqn}
and define $\Corrp{}{\mp}$ as
\begin{\eqn}
    \Corr{}{\mp}{p \vs q} \equiv \frac{1}{\pi} \IntInf{d\ppz} 
        \frac{\SpecG{}{}{\pp \vs q}}{\ppz-\pz\mp\ieps}.
    \label{eqn:CorrMP}
\end{\eqn}
Then the time-ordered and retarded correlations can be written as 
\begin{eqnarray}
    \Corr{}{T}{p \vs q} & = & 
    \LRcase{\Corr{}{-}{p \vs q}}_{\tworow{a=1/2}{b=0\ \ }}
        + \LRcase{\Corr{}{+}{p \vs q}}_{\tworow{a=0\ \ }{b=1/2}}, 
    \label{eqn:RLCcorrT}
    \\
    \Corr{}{R}{p \vs q} & = & \LRcase{\Corr{}{-}{p \vs q}}_{a=b=1/2.} 
    \label{eqn:RLCcorrR}
\end{eqnarray}

Let us first consider the spinor structure of 
the spectral function (\ref{eqn:DefSpecG}).
$\SpecGp{}{}$ has a $4\times4$ spinor structure 
and can be expanded in terms of a complete set of Dirac matrices.
The Lorentz covariance restricts the general form of $\SpecGp{}{}$ as
\begin{eqnarray}
    \SpecG{}{}{p \vs q} & = & \SpecGp{S}{} + \SpecGp{P}{}\gmf 
        + \SpecGp{\Vo}{}\sla{p} + \SpecGp{\Vt}{}\sla{q} 
        + \SpecGp{\Ao}{}\sla{p}\gmf + \SpecGp{\At}{}\sla{q}\gmf 
        \nonumber \\ & & 
        + \SpecGp{\To}{}(i\sg_{\mu\nu}p^{\mu}q^{\nu})
        + \SpecGp{\Tt}{}(\sgF_{\mu\nu}p^{\mu}q^{\nu}),
    \label{eqn:Dorig}
\end{eqnarray}
where we define the coefficients 
$\SpecGp{l}{}=\SpecG{l}{}{p^{2}, p \ip q, q^{2}}$ 
for $l=S,P,\Vo,\Vt,\Ao,\At,\To,\Tt$ and 
$\sgF_{\mu\nu} \equiv 
    \frac{1}{2}\varepsilon_{\mu\nu\alpha\beta}\,\sg^{\alpha\beta}
    =-i\sg_{\mu\nu}\gmf$ 
with a convention $\varepsilon^{0123}=1$.

Parity and time-reversal properties further restrict 
the spinor structure of $\SpecGp{}{}$.
Since the baryonic currents $\curr{A}{x}$ and $\curr{B}{x}$ have 
the same transformation properties with elementary Dirac fields 
under parity ($\Parity$) and time-reversal ($\TimeRev$) transformations, 
we have 
\begin{\eqn}
    \Parity\,\curr{A}{x}\Parity^{-1}=P\curr{A}{\xT},\;\;
    \Parity\,\currb{A}{x}\Parity^{-1}=\currb{A}{\xT}P^{-1},
\end{\eqn}
\begin{\eqn}
    \TimeRev\,\curr{A}{x}\TimeRev^{-1}=T\curr{A}{-\xT},\;\;
    \TimeRev\,\currb{A}{x}\TimeRev^{-1}=\currb{A}{-\xT}T^{-1},
\end{\eqn}
where 
$x^{\mu}=(\zero{x},\vect{x})$ and $\xT^{\mu} \equiv (\zero{x},-\vect{x})$.
$P$ ($T$) is a $4\times4$ matrix in the spinor space for 
the parity (time-reversal) transformation.
We assume the same transformation matrices $P$ and $T$ for $\curr{B}{x}$.
Under the Hermitian conjugate, $P$ and $T$, 
the Dirac matrices transform as 
\begin{eqnarray}
    \!\!\!\!\!\!\!\!\!\!\!\!
    \zero{\gm}\LRbkL{\LRbkM{1, \gmf,\, \gm_{\mu},\, \gm_{\mu}\gmf,\, 
        i\sg_{\mu\nu},\, \sgF_{\mu\nu}}}^{\dagger}\zero{\gm}
    & = & 
    \LRbkM{1,\, -\gmf,\, \gm_{\mu},\, \gm_{\mu}\gmf,\, 
        -i\sg_{\mu\nu},\, \sgF_{\mu\nu}}, 
    \label{eqn:HC}
    \\
    \!\!\!\!\!\!\!\!\!\!\!\!
    P \LRbkM{1, \gmf,\, \gm_{\mu},\, \gm_{\mu}\gmf,\,
        i\sg_{\mu\nu},\, \sgF_{\mu\nu}} P^{-1}
    & = & 
    \LRbkM{1,\, -\gmf,\, \gmT_{\mu},\, -\gmT_{\mu}\gmf,\, 
        i\sgT_{\mu\nu},\, -\sgFT_{\mu\nu}}, 
    \label{eqn:Parity}
    \\
    \!\!\!\!\!\!\!\!\!\!\!\!
    \LRbkL{\,T \LRbkM{1, \gmf,\, \gm_{\mu},\, \gm_{\mu}\gmf,\, 
        i\sg_{\mu\nu},\, \sgF_{\mu\nu}} T^{-1}}^{*}
    & = & 
    \LRbkM{1,\, \gmf,\, \gmT_{\mu},\, \gmT_{\mu}\gmf,\, 
        i\sgT_{\mu\nu},\, \sgFT_{\mu\nu}},
    \label{eqn:TimeRev}
\end{eqnarray}
where 
$\gm^{\mu}=(\zero{\gm},\vect{\gm})$, 
$\gmT^{\mu}\equiv(\zero{\gm},-\vect{\gm})$, 
$\sgT_{\mu\nu}\equiv \frac{i}{2}[\gmT_{\mu},\gmT_{\nu}]$ and 
$\sgFT_{\mu\nu} \equiv -i\sgT_{\mu\nu}\gmf$.
The state vector $\ket{q}$ is assumed to have the property, 
\begin{\eqn}
    \Parity\ket{q}=\ket{\qT},\;\; \TimeRev\ket{q}=\ket{\qT},
\end{\eqn}
with $q^{\mu}=(\zero{q},\vect{q})$ and 
$\qT^{\mu} \equiv (\zero{q},-\vect{q})$.

Because of the transformation properties of the baryonic current 
and the state vector shown above, 
the spectral function satisfies the following relations 
\begin{eqnarray}
    \LRcase{\SpecG{}{}{p \vs q}}_{\exchange{A}{B}} & = & 
    \zero{\gm}\LRbkM{\SpecG{}{}{p \vs q}}^{\dagger}\zero{\gm}, 
    \nonumber
    \\
    \SpecG{}{}{p \vs q} & = & P\,\SpecG{}{}{\pT \vs \qT}P^{-1}, 
    \label{eqn:SpecPT}
    \\
    \SpecG{}{}{p \vs q} & = & \LRbkM{T\,\SpecG{}{}{\pT \vs \qT}T^{-1}}^{*},
    \nonumber
\end{eqnarray}
where $\exchange{A\!}{\!B}$ stands for the exchange of $A$ and $B$.

Owing to Eq.(\ref{eqn:SpecPT}) and Eqs.(\ref{eqn:HC}-\ref{eqn:TimeRev}),
eight independent functions in Eq.(\ref{eqn:Dorig}) 
reduces to four functions such as 
\begin{\eqn}
    \SpecG{}{}{p \vs q} = \SpecGp{S}{}
        + \SpecGp{\Vo}{}\sla{p} + \SpecGp{\Vt}{}\sla{q}
        + \SpecGp{\To}{}(i\sg_{\mu\nu}p^{\mu}q^{\nu}),
    \label{eqn:SpecGF}
\end{\eqn}
where 
\begin{\eqn}
    \begin{array}{c}
        \begin{array}{cc}
        \LRcase{\SpecGp{l}{}\,}_{\exchange{A}{B}} = \SpecGp{l}{}
        \;\; \LRbk{l=S,\Vo,\Vt}, & 
        \LRcase{\SpecGp{\To}{}}_{\exchange{A}{B}} =-\SpecGp{\To}{},
        \end{array}
        \\
        \SpecGp{l}{}{}^{*} = \SpecGp{l}{} \;\; \LRbk{l=S,\Vo,\Vt,\To}.
    \end{array}
    \label{eqn:SpecGProp}
\end{\eqn}
Eqs.(\ref{eqn:SpecGF}) and (\ref{eqn:SpecGProp}) also imply that 
the correlation function $\Corrp{}{\mp}$ defined in Eq.(\ref{eqn:CorrMP}) 
has a form 
\begin{\eqn}
\Corr{}{\mp}{p \vs q}=
    \Corrp{S}{\mp} + \Corrp{\Vo}{\mp}\sla{p} + \Corrp{\Vt}{\mp}\sla{q} 
    + \Corrp{\To}{\mp}(i\sg_{\mu\nu}p^{\mu}q^{\nu})
\label{eqn:CorrGF0}
\end{\eqn}
where 
$\Corrp{l}{\mp}=\Corr{l}{\mp}{p^{2}, p\ip q, q^{2}}$ for $l=S,\Vo,\Vt,\To$
and 
\begin{\eqn}
    \LRcase{\Corrp{l}{\mp}}_{\exchange{A}{B}} = \Corrp{l}{\mp} 
        \;\; \LRbk{l=S,\Vo,\Vt}, \;\;
    \LRcase{\Corrp{\To}{\mp}}_{\exchange{A}{B}} = - \Corrp{\To}{\mp}.
    \label{eqn:CorrGF}
\end{\eqn}
Because of Eqs.(\ref{eqn:RLCcorrT}) and (\ref{eqn:RLCcorrR}), 
the time-ordered and retarded correlation functions have 
the same decomposition as Eq.(\ref{eqn:CorrGF0}).

Note that, for $A=B$, our results are fully consistent with 
the previous analysis in Ref.\cite{RF}.
In particular, the tensor terms 
$\SpecGp{\To}{}$ and $\Corrp{\To}{\mp}$ vanish in this case.
For $A \neq B$, our results are new.
The tensor terms do not vanish in this case unless 
$p^{\mu}$ and $q^{\mu}$ satisfy special conditions.

\subsection{Dispersion relations}

Eq.(\ref{eqn:CorrGF0}) enables us to decompose 
the dispersion relation (\ref{eqn:CorrMP}) 
into independent structures.
In this subsection we will work
in the rest frame of the state vector $\ket{q}$ 
$\LRbk{q^{\mu}=(\zero{q},\vzero)}$,
since it is sufficient for later applications.
For notational simplicity, 
we will omit the argument $q$ in $\Corrp{}{\mp}$, 
whenever we consider the rest frame of $\ket{q}$.
 
Under this simplification, Eq.(\ref{eqn:CorrGF0}) becomes 
\begin{\eqn}
    \Corr{}{\mp}{p}=
        \Corrp{S}{\mp} + \Corrp{V}{\mp}\zero{\gm} 
        - \LRbk{\Corrp{\Vo}{\mp} - \Corrp{\To}{\mp}\zero{q}\zero{\gm}} 
        (\vect{p} \ip \vect{\gm}), 
        \label{eqn:CorrMPp}
\end{\eqn}
where we have introduced
\begin{\eqn}
    \Corrp{V}{\mp} \equiv 
        \Corrp{\Vo}{\mp}\,\zero{p} + \Corrp{\Vt}{\mp}\,\zero{q}.
\end{\eqn}
Then the dispersion relation for $l=S,V$ reads 
\begin{\eqn}
    \Re \Corr{l}{\mp}{\omg,\va{p}} = \dispst \pm \frac{1}{\pi} 
        \PIntInf{d\omgp} \frac{\Im \Corr{l}{\mp}{\omgp,\va{p}}}{\omgp-\omg},
    \label{eqn:DispRel}
\end{\eqn}
where 
\begin{\eqn}
    \Im \Corr{l}{\mp}{\omg,\vect{p}} = \pm \frac{1}{2i} \limzp 
        \left[\Re \Corr{l}{\mp}{\omg + \ieps,\va{p}}
            - \Re \Corr{l}{\mp}{\omg - \ieps,\va{p}}\right],
    \label{eqn:DispRelInv}
\end{\eqn}
with $\omg \equiv \zero{p}$ 
and P stands for the principal value integral.
As far as $\vect{p} \neq \vzero$, 
the same dispersion relation holds for $l=\Vo,\To$.

Next let us decompose the correlation functions to even and odd parts 
under the transformation $\omg \leftrightarrow - \omg$ \cite{KM}:
\begin{\eqn}
    \Corr{l}{\mp}{\omg,\va{p}}=\CorrE{l\,}{\mp}{\omgsqr,\va{p}} 
        + \omg\,\CorrO{l\,}{\mp}{\omgsqr,\va{p}}.
\end{\eqn}
Then Eq.(\ref{eqn:DispRel}) reduces to a formula 
which relates the even (odd) part of $\Re \Corrp{}{\mp}$ with 
the odd (even) part of $\Im \Corrp{}{\mp}$: 
\begin{eqnarray}
    \Re \CorrE{l\,}{\mp}{s,\va{p}} & = & \pm \frac{1}{\pi} \PIntZI{d\spr}
        \frac{\, \Im \CorrO{l\,}{\mp}{\spr,\va{p}} \sqrt{\spr} \,}{\spr-s}{}, 
    \label{eqn:DispRelE}
    \\
    \Re \CorrO{l\,}{\mp}{s,\va{p}} & = & \pm \frac{1}{\pi} \PIntZI{d\spr}
        \frac{\, \Im \CorrE{l\,}{\mp}{\spr,\va{p}} / \sqrt{\spr} \,}{\spr-s}{}.
    \label{eqn:DispRelO}
\end{eqnarray}
Also, Eq.(\ref{eqn:DispRelInv}) reduces to 
\begin{eqnarray}
    \!\!\!\!\!\!\!\!\!\!\!
    \Im \CorrO{l\,}{\mp}{s,\va{p}} \!&\!\! = \!\!&\! \pm \frac{1}{2i} \limzp
        \left[\Re \CorrE{l\,}{\mp}{s \!+\! \ieps,\va{p}} 
            - \Re \CorrE{l\,}{\mp}{s \!-\! \ieps,\va{p}}\right] / \sqrt{s} {},
    \label{eqn:DispRelEInv}
    \\
    \!\!\!\!\!\!\!\!\!\!\!
    \Im \CorrE{l\,}{\mp}{s,\va{p}} \!&\!\! = \!\!&\! \pm \frac{1}{2i} \limzp
        \left[\Re \CorrO{l\,}{\mp}{s \!+\! \ieps,\va{p}} 
            - \Re \CorrO{l\,}{\mp}{s \!-\! \ieps,\va{p}}\right] \sqrt{s} {},
    \label{eqn:DispRelOInv}
\end{eqnarray}
for $l=S,V,\Vo,\To$ where $s \equiv \omgsqr$.

\subsection{Dispersion relations between 
Re${\bf\Corrp{}{T}}$ and Im${\bf\Corrp{}{R}}$}

The retarded correlation function defined in Eq.(\ref{eqn:RLCcorrR}) 
satisfies the same dispersion relations 
(\ref{eqn:DispRel}), (\ref{eqn:DispRelE}) and (\ref{eqn:DispRelO}).
Also, Eqs.(\ref{eqn:RLCcorrT}) and (\ref{eqn:SpecGProp}) 
imply that the real part of $\Corrp{}{T}$ 
and that of $\Corrp{}{R}$ are equal for each spinor component:
\begin{\eqn}
    \Re \Corrp{l}{T} = \Re \Corrp{l}{R}
        \;\;\; \LRbk{l=S,\Vo,\Vt,\To,V}.
    \label{eqn:ReTR}
\end{\eqn}
Therefore, the dispersion relations 
in the rest frame of the state vector $\ket{q}$ reads 
\begin{\eqn}
    \Re \Corr{l}{T}{\omg,\va{p}} = \dispst \frac{1}{\pi} 
        \PIntInf{d\omgp} \frac{\Im \Corr{l}{R}{\omgp,\va{p}}}{\omgp-\omg},
    \label{eqn:TR-DispRel}
\end{\eqn}
\begin{\eqn}
    \Im \Corr{l}{R}{\omg,\vect{p}} = \frac{1}{2i} \limzp 
        \left[\Re \Corr{l}{T}{\omg + \ieps,\va{p}}
            - \Re \Corr{l}{T}{\omg - \ieps,\va{p}}\right].
\label{eqn:TR-DispRelInv}
\end{\eqn}
By decomposing the above to the even and odd parts, one finds 
\begin{eqnarray}
    \Re \CorrE{l\,}{T}{s,\va{p}} \!&\!\! = \!\!&\! 
        \frac{1}{\pi} \PIntZI{d\spr}
        \frac{\, \Im \CorrO{l\,}{R}{\spr,\va{p}} \, \sqrt{\spr} \,}{\spr-s}{},
    \label{eqn:TR-DispRelE}
    \\
    \Re \CorrO{l\,}{T}{s,\va{p}} \!&\!\! = \!\!&\! 
        \frac{1}{\pi} \PIntZI{d\spr}
        \frac{\, \Im \CorrE{l\,}{R}{\spr,\va{p}} / \sqrt{\spr} \,}{\spr-s}{},
    \label{eqn:TR-DispRelO}
\end{eqnarray}
and 
\begin{eqnarray}
    \!\!\!\!\!\!\!\!\!\!\!
    \Im \CorrO{l\,}{R}{s,\va{p}} \!&\!\! = \!\!&\! \frac{1}{2i} \limzp
        \left[\Re \CorrE{l\,}{T}{s \!+\! \ieps,\va{p}} 
            - \Re \CorrE{l\,}{T}{s \!-\! \ieps,\va{p}}\right] / \sqrt{s} {},
    \label{eqn:TR-DispRelEInv}
    \\
    \!\!\!\!\!\!\!\!\!\!\!
    \Im \CorrE{l\,}{R}{s,\va{p}} \!&\!\! = \!\!&\! \frac{1}{2i} \limzp
        \left[\Re \CorrO{l\,}{T}{s \!+\! \ieps,\va{p}} 
            - \Re \CorrO{l\,}{T}{s \!-\! \ieps,\va{p}}\right] \sqrt{s} {},
    \label{eqn:TR-DispRelOInv}
\end{eqnarray}
$l=S,V,\Vo,\To$ with 
$\Corrp{V}{T,\, R} \equiv 
    \Corrp{\Vo}{T,\, R}\,\zero{p} + \Corrp{\Vt}{T,\, R}\,\zero{q}$.

In the sum rule analysis in later sections, 
$\Re \Corrp{l}{T}$ is evaluated by the operator product expansion, 
while phenomenological ansatzes are made for $\Im \Corrp{l}{R}$.

\section{STRUCTURE OF BARYON MIXING}
\setcounter{equation}{0}
\label{sec:MixForm}

In this section, we discuss the general structure of 
the baryon mixing.

\subsection{Definition of the mixing angle}

Let us consider a mass matrix for particles $A$ and $B$ 
which have definite quantum numbers but have different masses, 
\begin{\eqn}
    M = \LRbk{
    \begin{array}{c}
        \bra{A} \\ \bra{B}
    \end{array}
    }
    H
    \LRbk{
    \begin{array}{cc}
        \ket{A} & \ket{B}
    \end{array}
    }
    = \LRbk{
    \begin{array}{cc}
        \M_{A} & \EVt{A}{H}{B} \\ 
        \EVt{A}{H}{B} & \M_{B}
    \end{array}
    },
\end{\eqn}
where $H$ is the Hamiltonian of the system 
and $\ket{A}$ and $\ket{B}$ are one-particle states normalized 
as $\EV{A|A} = \EV{B|B} = 1$.
The diagonal matrix element is equal to the `mass' of the symmetric state, 
namely $\M_{A}=\EVt{A}{H}{A}$ and $\M_{B}=\EVt{B}{H}{B}$.
We choose the relative phase of the states $\ket{A}$ and $\ket{B}$ 
to be $\EVt{B}{H}{A}=\EVt{B}{H}{A}^{*}=\EVt{A}{H}{B}$.
The physical states $\ket{A}_{\Phys}$ and $\ket{B}_{\Phys}$ 
are represented as a linear combination of 
$\ket{A}$ and $\ket{B}$ by using 
the mixing angle $\AngleP$, 
\begin{eqnarray}
    \ket{A}_{\Phys} \!\! & = & \!\!
    \ket{A} \cos\AngleP + \ket{B} \sin\AngleP,
    \\
    \ket{B}_{\Phys} \!\! & = & \!\!
    \ket{B} \cos\AngleP - \ket{A} \sin\AngleP.
\end{eqnarray}
The mass matrix in terms of the physical states reads 
\begin{eqnarray}
    \!\!\!\! M_{\Phys} \!\!\!\! & = & \!\!\!\! 
    \left(
    \begin{array}{cc}
        \Ma - \DM^{\prime} / 2 &
        \EVt{A}{H}{B} \cos 2\AngleP + \DM \sin2\AngleP / 2 \\
        \EVt{A}{H}{B} \cos 2\AngleP + \DM \sin2\AngleP / 2 &
        \Ma + \DM^{\prime} / 2
    \end{array}
    \right)
\end{eqnarray}
where $\Ma\!=\!(\M_{A}+\M_{B})/2$, $\DM\!=\!\M_{B}-\M_{A}$ and 
$\DM^{\prime}\!= \DM \cos 2\AngleP - 2 \EVt{A}{H}{B} \sin 2\AngleP$.
Thus, for the weak mixing, $\AngleP$ is written as 
\begin{\eqn}
    \AngleP \simeq - \frac{\EVt{A}{H}{B}}{\DM}
    = - \frac{\EVt{A}{\Hint}{B}}{\DM}.
    \label{eqn:AngleME}
\end{\eqn}
where $\Hint$ is a part of $H$ 
which mixes the states $\ket{A}$ and $\ket{B}$.
When we consider baryon mixings, 
$\AngleP$ acquires spinor structures 
as will be discussed in the next subsection.

\subsection{Phenomenological ansatz}
\label{sec:PhenAnsatz}

Let us consider the case that $A$ and $B$ are baryons 
with different flavor quantum-numbers and different masses.
We further assume that the interaction which induces 
the mixing is small and can be treated in the 1st order perturbation.
Then the correlation function Eq.(\ref{eqn:corrabMT}) 
near the mass-shell of baryons $A$ and $B$ may be written as
\begin{\eqn}
    \Corr{}{T}{p \vs q}=\csop{A}\csop{B}
        \frac{1}{\sla{p}-\M_{A} + \ieps} \LRbk{\MMEM \!\,\, \DM} 
        \frac{1}{\sla{p}-\M_{B} + \ieps} .
\label{eqn:CorrTPole}
\end{\eqn}
where we omit the single pole contributions.
Here $\csop{A}$ is a coupling strength defined by 
$\left\langle 0 \left| \curr{A}{0} \right| A(\vect{p},s) \right\rangle$ 
$=$ $\csop{A}u_{A}(\vect{p},s)$, 
where 
$\overline{u}_{A}(\vect{p},r) \, u_{A}(\vect{p},s) = 2\M_{A} \delta_{rs}$ 
and $\ket{A(\vect{p},s)}$ satisfies the covariant normalization.
The phase of the state 
$\ket{A(\vect{p},s)}$ ($\ket{B(\vect{p},s)}$) 
is chosen so that $\csop{A}$ ($\csop{B}$) becomes real.
$\csop{A}$, $\csop{B}$, $\M_{A}$, $\M_{B}$ and $\DM=\M_{B}-\M_{A}$
take their unperturbed value 
in the 1st order of the mixing parameter $\MMEM$.
A diagrammatic illustration of Eq.(\ref{eqn:CorrTPole}) is shown 
in Fig.\ref{fig:Phendia}.
The state vector $\ket{q}$, 
which will be later identified with the isospin-asymmetric nuclear medium, 
is the major physical source of the mixing. 
In general, $\MMEM$ is a $4 \times 4$ matrix in the spinor space.
According to the discussion in Sec.\ref{sec:GPCorr}, 
the correlation function (\ref{eqn:CorrTPole}) 
must have a form 
\begin{\eqn}
    \Corr{}{T}{p \vs q} = 
        \Corrp{S}{T} + \Corrp{\Vo}{T}\sla{p} + \Corrp{\Vt}{T}\sla{q} 
        + \Corrp{\To}{T}(i\sg_{\mu\nu}p^{\mu}q^{\nu}),
    \label{eqn:TR-CorrGF0}
\end{\eqn}
where $\Corrp{l}{T}=\Corr{l}{T}{p^{2}, p\ip q, q^{2}}$ and 
\begin{\eqn}
    \LRcase{\Corrp{l}{T}}_{\exchange{A}{B}} = \Corrp{l}{T} 
    \;\; \LRbk{l=S,\Vo,\Vt}, \;\;
    \LRcase{\Corrp{\To}{T}}_{\exchange{A}{B}} = - \Corrp{\To}{T}.
    \label{eqn:TR-CorrGF}
\end{\eqn}
Thus the Eq.(\ref{eqn:TR-CorrGF0}) restricts the structure of $\MMEM$.
In fact, it cannot contain 
$\gmf$, $\gm_{\mu}\gmf$ and $\sgF_{\mu\nu}$, 
and one obtains 
\begin{\eqn}
    \MMEM=\MME{S} + \MME{\Vo}\sla{p} + \MME{\Vt}\sla{q} 
        + \MME{\To}(i\sg_{\mu\nu}p^{\mu}q^{\nu}),
\end{\eqn}
where the parameters $\MME{l}$\ $(l=S, \Vo, \Vt, \To)$ are real.

\FigPhendia{htbp}{width=0.5\linewidth}

Now, let us consider an effective Lagrangian $\Lint$ 
which describes Eq.(\ref{eqn:CorrTPole}),
\begin{\eqn} 
    \Lint(x) = \LRbk{\baryb{A}(x)\,\MMEM\,\bary{B}(x) 
        + \baryb{B}(x)\,\MMEMb\,\bary{A}(x)} \DM .
\label{eqn:Lint}
\end{\eqn}
Here $\MMEMb=\zero{\gm}{\MMEM}^{\dagger}\zero{\gm}$ and 
$\bary{A}\,(\bary{B})$ is the field that 
describes the particle $A\,(B)$.
By the parity and time-reversal invariance of $\Lint$, 
$\bary{A}$ and $\bary{B}$ have the same transformation matrices $P$ and $T$
under the parity ($\Parity$) and time-reversal ($\TimeRev$) transformations.
This is the same constraint for 
the interpolating operators $\currn{A}$ and $\currn{B}$ 
in Sec.\ref{sec:GPCorrSS}.
Therefore, the fields $\bary{A}$ and $\bary{B}$ 
can simultaneously satisfy the relations 
$\left\langle 0 \left| \bary{A}(0) \right| A(\vect{p},s) \right\rangle$ 
$=$ $u_{A}(\vect{p},s)$ and 
$\left\langle 0 \left| \bary{B}(0) \right| B(\vect{p},s) \right\rangle$ 
$=$ $u_{B}(\vect{p},s)$ 
with the states $\ket{A(\vect{p},s)}$ and $\ket{B(\vect{p},s)}$ 
defined below Eq.(\ref{eqn:CorrTPole}).
Then, $\EVt{B}{\Hint}{A}=\EVt{A}{\Hint}{B}$ and 
$\EVt{\anti{B}}{\Hint}{\anti{A}}=\EVt{\anti{A}}{\Hint}{\anti{B}}$ hold.

We have implicitly assumed that 
the interaction does not contain any derivatives, 
which is equivalent to the assumption that 
$\MMEM$ depends only on $q^{\mu}$ and not on $p^{\mu}$.
Then, $\MMEM$ reduces to a simple form,
\begin{\eqn}
    \MMEM=\MME{S} + \MME{\Vt} \sla{q}
    \label{eqn:MME}
\end{\eqn}
For later convenience, 
we define a dimensionless parameter $\MME{V}$ as 
\begin{\eqn}
    \MME{V}=\MME{\Vt}\zero{q}.
\end{\eqn}

We can make physical interpretation of the mixing angles 
$\MME{S}$ and $\MME{V}$ as follows.
Consider the particle at rest 
in the rest frame of the medium ($\vect{p}=\vect{q}=\vzero$).
Then the off-diagonal matrix elements of the Hamiltonian 
$\Hint$ $\LRbk{= - \int d^{3}\!\vect{x}\, \Lint(t,\vect{x})}$ 
read
\begin{eqnarray}
    \EVt{A}{\Hint}{B} & = & - \LRbk{\MME{S} + \MME{V}} \DM, 
    \label{eqn:MMEP}
    \\
    \EVt{\anti{A}}{\Hint}{\anti{B}} & = & - \LRbk{\MME{S} - \MME{V}} \DM.
    \label{eqn:MMEA}
\end{eqnarray}
where $\ket{A}$ and $\ket{B}$ are normalized as $\EV{A|A} = \EV{B|B} = 1$.
Together with Eq.(\ref{eqn:AngleME}), 
we thus find that $\AngleP$ ($\AngleA$) defined below 
corresponds to the mixing angle in the particle (anti-particle) channel: 
\begin{eqnarray}
    \AngleP & = & \MME{S} + \MME{V},
    \label{eqn:AngleP}
    \\
    \AngleA & = & \MME{S} - \MME{V}.
    \label{eqn:AngleA}
\end{eqnarray}
$\MME{S}$ and $\MME{V}$ have formal analogy with 
the scalar and vector self-energies ($\Sigma^{S}$ and $\Sigma^{V}$) 
of the nucleon in the nuclear medium.
The nucleon and the anti-nucleon feel 
an optical potential $\Sigma^{S}+\Sigma^{V}$ 
and $\Sigma^{S}-\Sigma^{V}$ respectively \cite{SW}.

Let us rewrite the correlation functions 
in terms of the mixing angles $\MME{S}$ and $\MME{V}$ defined above.
Since $\Re \Corrp{l}{T}$ has a form deduced from 
Eq.(\ref{eqn:CorrTPole}) and Eq.(\ref{eqn:MME}), 
we obtain 
\begin{eqnarray}
    \Re \Corrp{S}{T} & = & \csop{A}\csop{B} 
        \LRbkL{\frac{\MME{S}\M_{A}+ \MME{\Vt}(p \ip q)}{\M_{A}^{2}-p^{2}} 
        -\frac{\MME{S}\M_{B}+ \MME{\Vt}(p \ip q)}{\M_{B}^{2}-p^{2}} },
    \\
    \Re \Corrp{\Vo}{T} & = & \csop{A}\csop{B} 
        \LRbk{\MME{S} + \MME{\Vt} \frac{(p \ip q)}{\Ma}} 
        \LRbkL{\frac{1}{\M_{A}^{2}-p^{2}} - \frac{1}{\M_{B}^{2}-p^{2}}},
    \\
    \Re \Corrp{\Vt}{T} & = & \csop{A}\csop{B} 
        \LRbk{\MME{\Vt}\frac{\DM}{2\Ma}} 
        \LRbkL{\frac{\M_{A}}{\M_{A}^{2}-p^{2}} 
        + \frac{\M_{B}}{\M_{B}^{2}-p^{2}}},
    \\
    \Re \Corrp{\To}{T} & = & \csop{A}\csop{B} 
        \LRbk{-\MME{\Vt}\frac{\DM}{2 \Ma}} 
        \LRbkL{\dispst \frac{1}{\M_{A}^{2}-p^{2}} 
        - \frac{1}{\M_{B}^{2}-p^{2}}}.
\end{eqnarray}
We mention here that $\Re \Corrp{l}{T}$ given 
above satisfy Eq.(\ref{eqn:TR-CorrGF}), 
since the mixing angles have the property 
\[
    \LRcase{\MME{S,V}}_{\exchange{A}{B}}=-\MME{S,V},
\]
which is confirmed by Eqs.(\ref{eqn:MMEP}) and (\ref{eqn:MMEA}).

For $p^{\mu}=(\omg,\vzero)$ and $q^{\mu}=(\zero{q},\vzero)$, 
the correlation function (\ref{eqn:TR-CorrGF0}) is simplified to 
\begin{\eqn}
    \Corr{}{T}{\omg} = \Corr{S}{T}{\omg} + \Corr{V}{T}{\omg}\zero{\gm},
\end{\eqn}
and its real parts become
\begin{eqnarray}
    \Re \Corr{S}{T}{\omg} & = & 
        \csop{A}\csop{B} \LRbkL{
        \frac{\MME{S} \M_{A} + \MME{V} \omg}{\M_{A}^{2}-\omg^{2}} 
        - \frac{\MME{S} \M_{B} + \MME{V} \omg}{\M_{B}^{2}-\omg^{2}} 
    }, 
    \\ 
    \Re \Corr{V}{T}{\omg} & = & 
        \csop{A}\csop{B} \LRbkL{
        \frac{\MME{S} \omg + \MME{V} \M_{A}}{\M_{A}^{2}-\omg^{2}} 
        - \frac{\MME{S} \omg + \MME{V} \M_{B}}{\M_{B}^{2}-\omg^{2}} 
    }.
\end{eqnarray}
We decompose 
$\Corr{S,V}{T}{\omg}$ further into even and odd parts,
\begin{\eqn}
    \Corr{l}{T}{\omg}=\CorrE{l\,}{T}{\omgsqr}+\omg\,\CorrO{l\,}{T}{\omgsqr}, 
\end{\eqn}
$l=S,V$.
Then we obtain a simple formula to be used later 
\begin{eqnarray}
    \Re \CorrE{l\,}{T}{s} & = & \csop{A}\csop{B} \, \MME{\,l} 
    \LRbkL{\frac{\M_{A}}{\M_{A}^{2}-s}-\frac{\M_{B}}{\M_{B}^{2}-s}}, 
    \label{eqn:CorrE-Phen}
    \\
    \Re \CorrO{l\,}{T}{s} & = & \csop{A}\csop{B} \, \MME{\,l} 
    \LRbkL{\frac{1}{\M_{A}^{2}-s}-\frac{1}{\M_{B}^{2}-s}}, 
    \label{eqn:CorrO-Phen}
\end{eqnarray}
where $s\equiv\omgsqr$ and $l=S,V$.

\setcounter{equation}{0}
\section{OPERATOR PRODUCT EXPANSION}
\label{sec:OPEinMed}

In this section, we carry out the operator product expansion (OPE) 
of $\Corrp{}{T}$ up to dimension 4 
and evaluate the in-medium matrix elements of local operators. 
As is well-known, 
the operators with Lorentz indices should be retained 
since they do not vanish in the medium \cite{HL}.
Furthermore, we need to keep not only the iso-scalar operators 
but also the iso-vector ones 
to take into account the isospin asymmetry in the medium.

\subsection{OPE for $\szl$ mixed correlation function}

Taking $A=\Lambda$ and $B=\Sigmaz$, 
the retarded correlation function (\ref{eqn:corrabMR}) reads
\begin{\eqn}
    \Corr{}{R}{p \vs q} = 
        i\fourier{4}\EVq{R[\curr{\Lambda}{x}\currb{\Sigmaz}{0}]},
    \label{eqn:corrSLMR}
\end{\eqn}
which satisfies the dispersion relations 
(\ref{eqn:TR-DispRel}), (\ref{eqn:TR-DispRelE}) and (\ref{eqn:TR-DispRelO}).
On the other hand, 
the time-ordered correlation function (\ref{eqn:corrabMT}) reads 
\begin{\eqn}
    \Corr{}{T}{p \vs q} = 
        i\fourier{4}\EVq{T[\curr{\Lambda}{x}\currb{\Sigmaz}{0}]}, 
    \label{eqn:corrSLMT}
\end{\eqn}
which is useful for making OPE.
$\ket{q}$ is the state vector corresponding to 
the nuclear medium with total four-momentum $q^{\mu}$.

For the interpolating operators $\curr{\Lambda}{x}$ and $\curr{\Sigmaz}{x}$, 
we adopt the Ioffe's current \cite{IOFFE},
\begin{\eqn}
    \curr{\psi_{1}\psi_{2}\psi_{3}}{x} = 
        \varepsilon_{abc}\LRbk{\psi^{a{\rm T}}_{1}\!(x) 
            \, C \, \gm_{\mu} \, \psi^{b}_{2}(x)} 
                \gmf\gm^{\mu} \psi^{c}_{3}(x) 
    \label{eqn:IoffeCurr}
\end{\eqn}
where $\psi(x)$ is the quark field with flavor $\psi$, 
$C$ denotes the charge conjugation matrix 
and $a,\, b,\,c$ are color indices.
This current is symmetric under the exchange of $\psi_{1}$ and $\psi_{2}$, 
i.e., 
$\curr{\psi_{1}\psi_{2}\psi_{3}}{x}=\curr{\psi_{2}\psi_{1}\psi_{3}}{x}$.
Thus, $\curr{\Lambda}{x}$ and $\curr{\Sigmaz}{x}$ may be written as 
\begin{\eqn}
    \curr{\Lambda}{x} = i\, \sqrt{\frac{2}{3}} 
        \LRbkM{\curr{usd}{x}-\curr{dsu}{x}},
    \ \ \ 
    \curr{\Sigmaz}{x} = i\, \sqrt{2}\, \curr{uds}{x}.
    \label{eqn:CurrSL}
\end{\eqn}
These are the same interpolating operators 
used in the analysis of $\szl$ mixing in the vacuum \cite{NYM, ZHY}.
Under the time-reversal, $\curr{\Lambda}{x}$ and $\curr{\Sigmaz}{x}$ 
transform in the same way as the quark field $\psi(x)$.

Using Eq.(\ref{eqn:CurrSL}) and the above mentioned exchange property, 
the mixed correlation (\ref{eqn:corrSLMT}) becomes 
\begin{\eqn}
    \Corr{}{T}{p \vs q} = \frac{2}{\sqrt{3}} 
        \left[\,
            i\!\fourier{4}\EVq{T[\curr{usd}{x}\currb{uds}{0}]} 
                - \LRbkM{u \leftrightarrow d}
        \,\right].
    \label{eqn:corrSLMTu-d}
\end{\eqn}
We carry out OPE of this correlation up to dimension 4 operators.
The quark masses are kept up to the 1st order.
Since Eq.(\ref{eqn:corrSLMTu-d}) is anti-symmetric 
under the exchange of $u$ and $d$, 
$u-d$ symmetric terms such as the gluon condensate 
$\EV{{\dispst \frac{\alpha_{\!s}}{\pi} G^{2}}}$ do not appear. 
All the diagrams contributing up to the order we consider 
are drawn in Fig.\ref{fig:OPEdia}.
The explicit forms of OPE may be summarized as follows: 
\begin{\eqn}
    \Corr{}{T}{p \vs q}=\Corrp{S}{T} + \Corrp{\Vo}{T}\sla{p} 
        + \Corrp{\Vt}{T}\sla{q} + \Corrp{\To}{T}(i\sg_{\mu\nu}p^{\mu}q^{\nu}).
\end{\eqn}
\begin{eqnarray}
    \Re \Corrp{S}{T} \!\!&\!\! = \!\!&\!\! \CoeffSL \left[
        \frac{1}{64\pi^{2}} \LRbk{\md-\muu} 
        p^{4} \log(-p^{2})
        - \frac{1}{8\pi^{2}} \EV{\qc{d}-\qc{u}} 
        p^{2} \log(-p^{2})
        \right. 
    \nonumber \\ \!\!&\!\! \!\!&\!\! \quad\quad 
        {} - \frac{1}{8\pi^{2}} 
        \left\{ (2\muu\!+\!\md)\EV{\qcv{u}{\sla{n}}} \frac{}{} \right. 
    \nonumber \\ & & \quad\quad \quad\; \left. 
        \left. \frac{}{} - (\muu\!+\!2\md)\EV{\qcv{d}{\sla{n}}} 
        + (\md\!-\!\muu)\EV{\qcv{s}{\sla{n}}} \right\} 
        (n \ip p)\log(-p^{2})
        \right],
    \\ 
    \Re \Corrp{\Vo}{T} \!\!&\!\! = \!\!&\!\! \CoeffSL \left[
        - \frac{1}{24\pi^{2}} 
        \EV{\qc{d}-\qc{u}} 
        (n \ip p) \log(-p^{2})
        \right. 
    \nonumber \\ \!\!&\!\! \!\!&\!\! \quad\quad 
        {} + \frac{1}{8\pi^{2}} 
        \left\{ \ms\EV{\qc{d}-\qc{u}} + \LRbk{\md-\muu}\EV{\qc{s}} 
        \!\frac{}{}\!\right\} 
        \log(-p^{2})
    \nonumber \\ \!\!&\!\! \!\!&\!\! \quad\quad 
        {} - \frac{1}{72\pi^{2}} \left\{ \LRbk{\md\EV{\qc{d}}-\muu\EV{\qc{u}}} 
        \frac{}{} \right. 
    \nonumber \\ \!\!&\!\! \!\!&\!\! \quad\quad \quad\;
        \left. \left. \frac{}{} 
        - 4\EV{\qcv{d}{(n \ip i\cd)\sla{n}}-\qcv{u}{(n \ip i\cd)\sla{n}}} 
        \right\} \left(\log(-p^{2})+\frac{2 (n \ip p)^{2}}{p^{2}}\right) 
        \right],
    \\
    \Re \Corrp{\Vt}{T} \!\!&\!\! = \!\!&\!\! \CoeffSL \left[
        \frac{5}{48\pi^{2}} 
        \EV{\qcv{d}{\sla{n}}-\qcv{u}{\sla{n}}} 
        p^{2} \log(-p^{2}) 
        \right.
    \nonumber \\ \!\!&\!\! \!\!&\!\! \quad\quad 
        {} + \frac{1}{18\pi^{2}} \left\{ \LRbk{\md\EV{\qc{d}}-\muu\EV{\qc{u}}} 
        \frac{}{} \right. 
    \nonumber \\ \!\!&\!\! \!\!&\!\! \quad\quad \quad\;
        \left. \left. \frac{}{} 
        {} - 4\EV{\qcv{d}{(n \ip i\cd)\sla{n}}-\qcv{u}{(n \ip i\cd)\sla{n}}} 
        \right\} (n \ip p)\log(-p^{2}) 
        \right] / \sqrt{q^{2}},
    \\
    \Re \Corrp{\To}{T} \!\!&\!\! = \!\!&\!\! \CoeffSL \left[
        \frac{1}{8\pi^{2}} \left\{
        \LRbk{\ms-\md}\EV{\qcv{u}{\sla{n}}}
            - \LRbk{\ms-\muu}\EV{\qcv{d}{\sla{n}}}
        \frac{}{} \right.\right. 
    \nonumber \\ 
    \!\!&\!\! \!\!&\!\! \quad\quad \quad\quad \left.\left. \frac{}{} 
        + \LRbk{\md-\muu}\EV{\qcv{s}{\sla{n}}}
        \right\} \log(-p^{2}) 
        \right] / \sqrt{q^{2}},
\end{eqnarray}
where we have defined a normal vector 
$n^{\mu} \equiv q^{\mu}/\sqrt{q^{2}}$ 
characterizing the nuclear medium.
Also we have replaced the in-medium matrix elements 
$\EVt{q}{\cdots}{q}$ by $\EV{\cdots}$ for simplicity.

\FigOPEdia{htbp}{scale=0.8}

The in-medium expectation values of isospin-asymmetric operators 
beyond dimension 4 have large uncertainties.
For example, $\EV{\qnv{d}{(\sg \ip G)}-\qnv{u}{(\sg \ip G)}}$, 
which may give a major contribution to the OPE at dimension 5, 
is not known.
In fact, even its isospin-symmetric partner $\EV{\qnv{q}{(\sg \ip G)}}$ 
has large error $(-0.33 \GeV^{2} \sim +0.66 \GeV^{2}) \ip \DensT$ 
with $\DensT$ being total nuclear medium density \cite{CFGJ}. 
Therefore, we limit ourselves to the isospin-asymmetric operators 
up to dimension 4 in this paper.

In the rest frame of the medium with $n^{\mu} = (1,\vzero)$, 
the decompositions to even and odd parts of $\Re \Corrp{S,\,V}{T}$ 
are written as 
\begin{eqnarray}
    \!\!\!\!\!
    \Re \CorrE{S}{T}{s,\va{p}} \!\!&\!\! = \!\!&\!\! 
        \LRbk{\frac{1}{16\pi^{4}}} \CoeffSL 
        \left[ 
            \frac{1}{4} \dmm\, p^{4} \log(-p^{2})
            + \frac{(-4\pi^{2})}{2} \EV{\diff{\qc{q}}} p^{2} \log(-p^{2}) 
        \right],
    \label{eqn:CorrSE-OPE}
    \\ 
    \!\!\!\!\!
    \Re \CorrE{V}{T}{s,\va{p}} \!\!&\!\! = \!\!&\!\! 
        \LRbk{\frac{1}{16\pi^{4}}} \CoeffSL 
        \left[
            \frac{(4\pi^{2})}{4} \EV{\diff{\qn{q}}} p^{2} \log(-p^{2}) 
        \right],
    \label{eqn:CorrVE-OPE}
    \\ 
    \!\!\!\!\!
    \Re \CorrO{S}{T}{s,\va{p}} \!\!&\!\! = \!\!&\!\! 
        \LRbk{\frac{1}{16\pi^{4}}} \CoeffSL 
        \left[ 
            \frac{(4\pi^{2})}{2} 
            \left\{ 
                \dmm \LRbk{\EV{\ave{\qn{q}}}-\EV{\qn{s}}} 
                - \ma \EV{\diff{\qn{q}}} 
                \!\frac{}{}\! 
            \right\} 
            \log(-p^{2}) 
        \right],
    \label{eqn:CorrSO-OPE}
    \nonumber \\ & & 
    \\ 
    \!\!\!\!\!
    \Re \CorrO{V}{T}{s,\va{p}} \!\!&\!\! = \!\!&\!\! 
        \LRbk{\frac{1}{16\pi^{4}}} \CoeffSL 
        \left[ \frac{(4\pi^{2})}{2} \left\{ 
        \left( \ms \EV{\diff{\qc{q}}} + \dmm \EV{\qc{s}} + X \right) 
        \log(-p^{2}) 
    \frac{}{} \right. \right. \nonumber \\ \!\!&\!\! \!\!&\!\! \left. \left. 
    \quad\quad\quad\quad \quad\quad\quad\quad 
    \quad\quad\quad\quad \quad\quad\quad\quad 
    \quad\quad\quad\quad
        - \frac{2}{3}\, X \!\LRbk{\frac{s}{p^{2}}}
        \right\} \right],
    \label{eqn:CorrVO-OPE}
\end{eqnarray}
with 
\begin{\eqn}
    X \equiv \frac{1}{3} \LRbkM{ 
        \EV{\diff{\qcv{q}{\,i\!\slacd}}} 
            - 4 \EV{\diff{\qcv{q}{\,i\zero{\cd}\zero{\gm}}}}
},
\end{\eqn}
and $p^{2}=s-\vect{p}^{2}$.
In the above formulas, $\widehat{\quad}$ denotes the $u-d$ average, namely 
$\ma=\LRbk{\muu+\md}/2$, $\ave{\qn{q}}=\LRbk{\qn{u}+\qn{d}}/2$ and 
$\ave{\qc{q}}=\LRbk{\qc{u}+\qc{d}}/2$, 
while $\delta$ denotes $d-u$ difference, namely 
$\dmm=\md-\muu$, $\diff{\qc{q}}=\qc{d}-\qc{u}$ and 
$\diff{\qn{q}}=\qn{d}-\qn{u}$.

\subsection{In-medium condensates}

In the previous subsection, we have encountered 
various $u-d$ symmetric and $u-d$ anti-symmetric condensates.
In this subsection, we will evaluate those 
in a model independent way using the low density expansion.

First of all, the expectation value of the local operator $\OpO$ has 
a vacuum part which is density independent and 
the medium part which is density dependent; 
\begin{\eqn}
    \EV{\OpO} = \EVV{\OpO} + \EVM{\OpO}.
\end{\eqn}
At low density, $\EVM{\OpO}$ is expanded as 
\begin{\eqn}
    \EVM{\OpO} = \EVs{\OpO}{p} \Dens{p} + \EVs{\OpO}{n} \Dens{n} + \cdots,
    \label{eqn:MedEV}
\end{\eqn}
where $\Dens{p}$ ($\Dens{n}$) is the proton (neutron) density, 
and $\EVs{\OpO}{p}$ ($\EVs{\OpO}{n}$) is the spin-averaged expectation value 
taken by the one particle state of the proton (neutron).
\begin{\eqn}
    \EVs{\OpO}{N} = \int\!d^{3}\!\vect{x} 
        \LRbkM{\EVt{N}{\OpO}{N}- \EVt{0}{\OpO}{0}},
\end{\eqn}
where $\EV{N|N}=1$ for $N=p,n$.

Since the vector condensate $\EV{\qn{q}}$ 
is nothing but the quark number density, we have 
\begin{\eqn}
    \begin{array}{ccc}
        \EV{\ave{\qn{q}}} = {\dispst \frac{3}{2}}\,\DensT, & 
        \EV{\diff{\qn{q}}} = \dDens, 
        \label{eqn:dqn} & 
        \EV{\qn{s}} = 0, 
    \end{array}
\end{\eqn}
where $\DensT \,(\equiv \Dens{p} + \Dens{n})$ is the total nucleon density, 
$\dDens \,(\equiv \Dens{n} - \Dens{p})$ is the $n-p$ asymmetry.

The scalar condensate 
$\EV{\qc{q}} = \EVV{\qc{q}} + \EVM{\qc{q}}$ 
is evaluated by the Feynman-Hellmann theorem \cite{DL};
\begin{\eqn}
    \EVM{\frac{\rd\HQCD}{\rd\lambda}} \!\! = 
        \frac{\rd {\mathcal E}}{\rd\lambda},
    \label{eqn:FHT}
\end{\eqn}
where $\lambda$ is a parameter in the QCD Hamiltonian, 
and ${\mathcal E}$ is the energy density of the nuclear medium.
At low density, 
\begin{\eqn}
    {\mathcal E} \equiv \EVM{\HQCD} 
        \simeq \M_{p}\,\Dens{p} + \M_{n}\,\Dens{n} 
        = \MaN \, \DensT + \dMN \, \dDens/2 
\end{\eqn}
where $\MaN=(\M_{p}+\M_{n})/2$ and $\dMN=\M_{n}-\M_{p}$.
Under the choice $\lambda=2\ma$ together with 
the mass term of the QCD Hamiltonian 
\begin{\eqn}
    \HQCD_{\rm mass} = \ma(\qc{u}+\qc{d}) + \frac{\dmm}{2}(\qc{d}-\qc{u}) 
        + \ms\qc{s},
\end{\eqn}
one finds from Eq.(\ref{eqn:FHT})
\begin{\eqn}
    \EVM{\ave{\qc{q}}} \simeq \frac{\rd \MaN}{2\,\rd\ma}\,\DensT 
        = \frac{\sg_{N}}{2\ma}\,\DensT ,
\end{\eqn}
where we have used a definition of the nucleon sigma-term 
$\sg_{N}=\ma \, {\dispst \frac{\rd \MaN}{\rd \ma}}$.
\\
On the other hand, the choice $\lambda=\dmm/2$, 
with Eq.(\ref{eqn:FHT}) gives 
\begin{\eqn}
    \EVM{\diff{\qc{q}}} \simeq \frac{\rd(\dMN)}{\rd(\dmm)}\,\dDens ,
    \label{eqn:dqcDens}
\end{\eqn}
which is valid up to the 1st order in $\dmm$.
\\
The strange-quark condensate up to the 1st order in $\DensT$ reads
\begin{\eqn}
    \EVM{\qc{s}} 
        \simeq \frac{1}{2} \LRbk{\EVs{\qc{s}}{p}+\EVs{\qc{s}}{n}} \DensT 
        \equiv \frac{y}{2} \LRbk{\EVs{\widehat{\qc{q}}}{p}
            + \EVs{\widehat{\qc{q}}}{n}} \DensT.
\end{\eqn}
Here $y$ is a parameter characterizing 
the OZI violation in the nucleon \cite{GL}.

Thus the quark condensates $\EV{\qc{q}}$ in isospin-asymmetric 
nuclear medium are summarized as follows: 
\begin{eqnarray}
    \EV{\ave{\qc{q}}} & \simeq & \EVV{\widehat{\qc{q}}} 
        + \frac{\sg_{N}}{2 \ma} \, \DensT ,
    \\ 
    \EV{\diff{\qc{q}}} & \simeq & \EVV{\diff{\qc{q}}} 
        + \frac{\rd(\dMN)}{\rd(\dmm)} \, \dDens ,
    \\ 
    \EV{\qc{s}} & \simeq & \EVV{\qc{s}} 
        + y\,\frac{\sg_{N}}{2\ma} \, \DensT .
\end{eqnarray}

By using the result of evaluation of 
$\EV{\diff{\qcv{q}{\,i\zero{\cd}\zero{\gm}}}}$ 
in Appendix, we obtain 
\begin{eqnarray}
    X & \equiv & \frac{1}{3} \LRbkM{
        \EV{\diff{\qcv{q}{\,i\!\slacd}}} 
            - 4 \EV{\diff{\qcv{q}{\,i\zero{\cd}\zero{\gm}}}}
        } \nonumber \\ 
    & \simeq & \frac{1}{2} \, \diff{A^{q}_{2}(\mu^{2})} 
    \LRbk{\MaN \, \dDens + \dMN \, \DensT/2},
\end{eqnarray}
where $A^{q}_{2}(\mu^{2})$ is the 2nd moment of 
the parton distribution function of the proton 
defined by Eq.(\ref{eqn:PDFP}) in Appendix.

\subsection{Summary of OPE for $\mathbf{Re \Corrp{}{T}}$}
\label{sec:OPESummary}

In the isospin-asymmetric nuclear medium, 
the OPE of the correlation function is finally expressed as
\begin{eqnarray}
    \!\!\!\!\!\!\!\!\!\!\!\!
    \Re \CorrE{S}{T}{s,\va{p}} \!\!&\!\!=\!\!&\!\! 
        \LRbk{\frac{1}{16\pi^{4}}} \CoeffSL 
        \left[ 
            \frac{1}{4} \dmm\, p^{4} \log(-p^{2})
            + \frac{(-4\pi^{2})}{2} \EV{\diff{\qc{q}}} p^{2} \log(-p^{2}) 
        \right], 
    \label{eqn:CorrSE-OPEM}
    \\ 
    \!\!\!\!\!\!\!\!\!\!\!\!
    \Re \CorrE{V}{T}{s,\va{p}} \!\!&\!\!=\!\!&\!\! 
        \LRbk{\frac{1}{16\pi^{4}}} \CoeffSL 
        \left[ 
            \frac{(4\pi^{2})}{4} \EV{\diff{\qn{q}}} p^{2} \log(-p^{2}) 
        \right], 
    \label{eqn:CorrVE-OPEM}
    \\ 
    \!\!\!\!\!\!\!\!\!\!\!\!
    \Re \CorrO{S}{T}{s,\va{p}} \!\!&\!\!=\!\!&\!\! 
        \,\order (\ma \, \dDens ,\, \dmm \, \DensT),
    \label{eqn:CorrSO-OPEM}
    \\ 
    \!\!\!\!\!\!\!\!\!\!\!\!
    \Re \CorrO{V}{T}{s,\va{p}} \!\!&\!\!=\!\!&\!\! 
        \LRbk{\frac{1}{16\pi^{4}}} \CoeffSL 
        \left[ \frac{(4\pi^{2})}{2} \left\{
        \left( \ms \EVV{\diff{\qc{q}}} + \dmm \EVV{\qc{s}} \right) 
        \log(-p^{2}) 
        \frac{}{} \right. \right. \nonumber \\ 
    \!\!\!\!\!\!\!\!\!\!\!\!
    \!\!&\!\! \!\!&\!\! \quad\quad \left. \left. \!
        {} + \frac{1}{2} \, \diff{A^{q}_{2}(\mu^{2})} 
        \left( \MaN \, \dDens + \dMN \, \DensT/2 \right) 
        \left( \log(-p^{2})-\frac{2}{3} \, \frac{s}{p^{2}} \right) 
        \right\} \right]
    \nonumber \\ 
    \!\!\!\!\!\!\!\!\!\!\!\!
    \!\!&\!\! \!\!&\!\! \quad\quad
        {} + \order (\ms \, \dDens ,\, \dmm \, \DensT),
    \label{eqn:CorrVO-OPEM}
\end{eqnarray}
where $p^{2}=s-\vect{p}^{2}$. 

Above formulas are valid 
up to the 1st order in the quark masses and the baryon density.
In other words, the terms such as 
$\order(\dmm)$ and $\order(\dDens)$ are kept, 
while the terms such as 
$\order(m_{u,d,s} \, \DensT)$, $\order(m_{u,d,s} \, \dDens)$ are neglected.
Within this approximation, 
the scalar-odd correlation $\CorrOp{S}{T}$ vanishes as shown above 
and cannot be used to construct sum rules.
Since it is theoretically consistent to use the correlation functions 
with the same reflection symmetry under $\omega \leftrightarrow - \omega$, 
we will exclusively use the ``even'' correlations 
$\CorrEp{S}{T}$ and $\CorrEp{V}{T}$ in the following analyses.

\setcounter{equation}{0}
\section{QCD SUM RULES}
\label{sec:QSR}

In this section, we construct 
finite energy sum rules (FESR) \cite{KPT} 
and Borel sum rules (BSR) \cite{SVZ} 
on the basis of the retarded correlation function (\ref{eqn:corrSLMR}) and 
the dispersion relations (\ref{eqn:TR-DispRelE}) and (\ref{eqn:TR-DispRelO}).
For the phenomenological side, we use the ansatz given in 
Eqs.(\ref{eqn:CorrE-Phen}) and (\ref{eqn:CorrO-Phen}).
The OPE side is given in 
Eqs.(\ref{eqn:CorrSE-OPE}-\ref{eqn:CorrVO-OPE}).

\subsection{Finite Energy Sum Rules (FESR)}

In FESR, we identify the integral of 
$\Im \Corrp{}{R}$ extracted from OPE with 
that introduced phenomenologically:
\begin{eqnarray}
    \int^{\ThrO{\,l\,}}_{0} \!\!\!\! ds \, 
        \LRbkM{\Im \CorrO{l}{\Phen}{s,\va{p}} \sqrt{s}} s^{n} & = & 
    \int^{\ThrO{\,l\,}}_{0} \!\!\!\! ds \, 
        \LRbkM{\Im \CorrO{l}{\OPE}{s,\va{p}} \sqrt{s}} s^{n}, 
    \label{eqn:FEsum-O} 
    \\ 
    \int^{\ThrE{\,l\,}}_{0} \!\!\!\! ds \, 
        \LRbkM{\Im \CorrE{l}{\Phen}{s,\va{p}} / \sqrt{s}} s^{n} & = & 
    \int^{\ThrE{\,l\,}}_{0} \!\!\!\! ds \, 
        \LRbkM{\Im \CorrE{l}{\OPE}{s,\va{p}} / \sqrt{s}} s^{n}, 
    \label{eqn:FEsum-E} 
\end{eqnarray}
where $s = \omega^{2}$.
$\ThrO{\,l \,}$ and $\ThrE{\,l \,}$ are the continuum thresholds of 
$\Im \CorrOp{l}{\Phen}$ and $\Im \CorrEp{l}{\Phen}$ respectively.
These phenomenological spectral functions at $\vect{p}=\vzero$ 
in the left hand side of the sum rule are obtained as follows.
We substitute the ansatz 
(\ref{eqn:CorrE-Phen}) and (\ref{eqn:CorrO-Phen}) 
with $A\!=\!\Lambda$, $B\!=\!\Sigmaz$ 
into Eqs.(\ref{eqn:TR-DispRelEInv}) and (\ref{eqn:TR-DispRelOInv}) 
to obtain 
\begin{eqnarray}
    \Im \Corr{S}{\Phen}{\omg} & = & 
        \frac{\pi}{2} \csop{\Lambda}\csop{\Sigmaz} \, 
        \left[ 
            \LRbk{\MME{S} + \MME{V}} 
            \left\{ 
                \, \delta \LRbk{\omg - \ML} - \delta \LRbk{\omg - \MSz} 
            \right\} 
    \right. \nonumber \\ & & \left. \quad\quad\quad\quad 
            - \LRbk{\MME{S} - \MME{V}} 
            \left\{ 
                \, \delta \LRbk{\omg + \ML} - \delta \LRbk{\omg + \MSz} 
            \right\} 
        \right], 
    \label{eqn:ImCorrS-PhenM}
    \\
    \Im \Corr{V}{\Phen}{\omg} & = & 
        \frac{\pi}{2} \csop{\Lambda}\csop{\Sigmaz} \, 
        \left[ 
            \LRbk{\MME{S} + \MME{V}} 
            \left\{ 
                \, \delta \LRbk{\omg - \ML} - \delta \LRbk{\omg - \MSz} 
            \right\} 
    \right. \nonumber \\ & & \left. \quad\quad\quad\quad 
            + \LRbk{\MME{S} - \MME{V}} 
            \left\{ 
                \, \delta \LRbk{\omg + \ML} - \delta \LRbk{\omg + \MSz} 
            \right\} 
        \right]. 
    \label{eqn:ImCorrV-PhenM}
\end{eqnarray}
The even-odd decompositions of the above formula give 
$\Im \CorrOp{l}{\Phen}$ and $\Im \CorrEp{l}{\Phen}$.
As we have discussed 
in Sec.\ref{sec:PhenAnsatz} and Sec.\ref{sec:OPESummary}, 
we consider only the first-order effect of the isospin-asymmetry 
on the mixed correlation function 
and neglect the effects of 
$\order(m_{u,d,s} \, \DensT)$ and $\order(m_{u,d,s} \, \dDens)$.
Therefore, the pole positions $\ML$ and $\MSz$ take their vacuum value 
and only the pole residues are affected in a different way 
in even and odd spectral functions.
The situation is the also same for excited states of $\Lambda$ and $\Sigmaz$.
Therefore, in the present approximation, the continuum threshold 
also takes their vacuum value ($\Thr = \ThrE{\,l\,} = \ThrO{\,l\,}$) 
and only the height of the continuum is affected by the isospin asymmetry.

The OPE motivated spectral functions at $\vect{p}=\vzero$ 
in the right hand side of the sum rule are obtained as follows.
We substitute Eqs.(\ref{eqn:CorrSE-OPE}-\ref{eqn:CorrVO-OPE}) 
together with Eq.(\ref{eqn:ReTR}) into 
Eqs.(\ref{eqn:TR-DispRelEInv}) and (\ref{eqn:TR-DispRelOInv}) 
to obtain, 
\begin{eqnarray}
    \!\!\!\!\!\!\!\!\!\!\!\! \!\!&\!\! \!\!&\!\!
    \Im \Corr{S}{\OPE}{\omg} = -\pi \, \sgn{\omg} 
        \LRbk{\frac{1}{16\pi^{4}}} \CoeffSL 
        \left[ \,\frac{1}{4} \dmm\, \omg^{4} 
        + \frac{(-4\pi^{2})}{2} \EV{\diff{\qc{q}}} \omg^{2} \right. 
    \nonumber \\ 
    \!\!\!\!\!\!\!\!\!\!\!\! \!\!&\!\! \!\!&\!\! 
    \quad \left. \frac{}{} 
        + \frac{(4\pi^{2})}{2} 
            \left\{ \dmm \left( \EV{\ave{\qn{q}}}-\EV{\qn{s}} \right) 
                - \ma \EV{\diff{\qn{q}}} \right\} \omg \right] ,
    \label{eqn:ImCorrS-OPEM}
    \\ 
    \!\!\!\!\!\!\!\!\!\!\!\! \!\!&\!\! \!\!&\!\!
    \Im \Corr{V}{\OPE}{\omg} = -\pi \, \sgn{\omg} 
        \LRbk{\frac{1}{16\pi^{4}}} \CoeffSL 
        \left[ \frac{(4\pi^{2})}{4} \EV{\diff{\qn{q}}} \omg^{2} \right.
    \nonumber \\ 
    \!\!\!\!\!\!\!\!\!\!\!\! \!\!&\!\! \!\!&\!\! 
    \quad \left. {}
        + \frac{(4\pi^{2})}{2} 
            \left\{ 
            \left( \ms \EV{\diff{\qc{q}}} + \dmm \EV{\qc{s}} \right) 
            + \frac{1}{2} \, \diff{A^{q}_{2}(\mu^{2})} 
                \left( \MaN \, \dDens + \dMN \, \DensT/2 \right) 
            \right\} \omg \right].
    \label{eqn:ImCorrV-OPEM}
\end{eqnarray}
Here, $\sgn{\omg}=\omg/\!\abs{\omg}\ (\omg \neq 0)$ and $\sgn{0}=0$.
The even-odd decompositions of the above formula give 
$\Im \CorrOp{l}{\rm OPE}$ and $\Im \CorrEp{l}{\rm OPE}$.
The isospin-asymmetric condensates affect 
the magnitude (height) of even and odd components 
in a different way.

As we have mentioned at the end of Sec.\ref{sec:OPEinMed}, 
we use $\Re \CorrEp{l}{T}$ for the actual analysis, 
which corresponds to adopt the sum rule (\ref{eqn:FEsum-O}). 
Remember that the even part $\Re \CorrEp{l}{T}$ is related to 
the odd part $\Im \CorrOp{l}{R}$ 
through the dispersion relation (\ref{eqn:TR-DispRelE}).
Resulting FESR for $\vect{p}=\vzero$ reads
\begin{eqnarray}
    \BetaSL \LRbk{\MSz^{2n+1}-\ML^{2n+1}} \MME{S} & = & 
    \CoeffSL \left[ 
        \, \frac{1}{4} \dmm\, \frac{\Thr^{n+3}}{n+3}
        + \frac{(-4\pi^{2})}{2} \EV{\diff{\qc{q}}} \frac{\Thr^{n+2}}{n+2} 
    \right], 
    \label{eqn:FESR-SE}
    \\ 
    \BetaSL \LRbk{\MSz^{2n+1}-\ML^{2n+1}} \MME{V} & = & 
    \CoeffSL \left[ 
        \frac{(4\pi^{2})}{4} \EV{\diff{\qn{q}}} \frac{\Thr^{n+2}}{n+2} 
    \right], 
    \label{eqn:FESR-VE}
\end{eqnarray}
where $\BetaSL=16\pi^{4}\csop{\Lambda}\csop{\Sigmaz}$.
As is evident from the right hand side of Eq.(\ref{eqn:FESR-VE}), 
$\MME{V}$ appears only in the nuclear medium, 
$\MME{V} = \MME{V}_{\Med}$.
On the other hand, $\MME{S}$ has both vacuum part and in-medium part 
$\MME{S} = \MME{}_{\Vac} + \MME{S}_{\Med}$. 
We subtract out the vacuum part from the sum rule 
(\ref{eqn:FESR-SE}) to obtain 
\begin{\eqn}
    \BetaSL \LRbk{\MSz^{2n+1}-\ML^{2n+1}} \MME{S}_{\Med} = 
    \CoeffSL \left[ 
        \frac{(-4\pi^{2})}{2} \EVs{\diff{\qc{q}}}{\Med} \frac{\Thr^{n+2}}{n+2}
    \right]. 
    \label{eqn:FESR-SEM}
\end{\eqn}
Combining Eqs.(\ref{eqn:FESR-VE}) and (\ref{eqn:FESR-SEM}), 
one has a simple formula for the ratio $\MME{V}_{\Med}/\MME{S}_{\Med}$:
\begin{\eqn}
    \frac{\MME{V}_{\Med}}{\MME{S}_{\Med}} = 
        - \frac{1}{2}\, \frac{\EV{\diff{\qn{q}}}}{\EVs{\diff{\qc{q}}}{\Med}}
    \simeq - \LRbkM{2\,\frac{\rd(\dMN)}{\rd(\dmm)}}^{-1},
    \label{eqn:FESRratio}
\end{\eqn}
where we have used Eqs.(\ref{eqn:dqn}) and (\ref{eqn:dqcDens}) 
for the last equality with $\dMN=\M_{n}-\M_{p}$ and $\dmm=\md-\muu$. 

Let us estimate the right hand side of Eq.(\ref{eqn:FESRratio}).
The $n-p$ mass difference $\dMN = 1.29 \,{\MeV}$ 
is known to be decomposed into two parts, 
$\dMN \simeq \dMN^{\EM} + \dMN^{\QCD}$. 
Here $\dMN^{\EM} (\simeq -0.76 \,{\MeV})$ originates from 
the electromagnetic interaction of $\order(\alpha)$, 
while $\dMN^{\QCD} (\simeq 2.04 \,{\MeV})$ is due to 
the $u-d$ quark mass difference of $\order(\dmm)$ \cite{GL}.
Therefore, in the leading order of $\dmm$ and $\alpha$, one finds
\begin{\eqn}
    \frac{\rd(\dMN)}{\rd(\dmm)} 
        = \frac{\rd(\dMN^{\QCD})}{\rd(\dmm)} 
        = \frac{\dMN^{\QCD}}{\dmm} = 0.52 \LRbk{\frac{3.9\,{\MeV}}{\dmm}},
    \label{eqn:Mm052}
\end{\eqn}
where we have used $\dmm = 3.9 \,{\MeV}$ 
as a typical value at the renormalization 
$\mu^{2} = 1 \,{\GeV}^{2}$ \cite{GL, NYM} 
(See also Table \ref{tb:QCDparameter}).
Thus we find that the scalar and vector mixing angles induced 
by the nuclear medium have opposite sign and approximately equal in magnitude, 
\begin{\eqn}
    \MME{V}_{\Med} / \MME{S}_{\Med} \sim - 1.
\end{\eqn}
This together with the definition of 
the total mixing angles Eqs.(\ref{eqn:AngleP}) and (\ref{eqn:AngleA}) implies 
that the medium modification of the particle mixing is 
largely cancelled between the scalar and vector, 
while the anti-particle mixing is enhanced in medium.
The magnitude of the mixing angles will be discussed in Sec.\ref{sec:NumA}.

\subsection{Borel sum rules (BSR)}

In BSR, we make a Borel transform of the dispersion relations 
(\ref{eqn:TR-DispRelE}) and (\ref{eqn:TR-DispRelO}) 
for the retarded correlation (\ref{eqn:corrSLMR}) 
in the deep Euclidian region $s=\omega^{2} \rightarrow - \infty$: 
\begin{eqnarray}
    \Borel\LRbkL{ \frac{1}{\pi} \PIntZI{d\spr} \, 
    \frac{\Im \CorrO{l}{\Phen}{\spr,\va{p}} \sqrt{\spr}}{\spr-s}} & = & 
    \Borel\LRbkL{\Re \CorrE{l}{\OPE}{s,\va{p}}}, 
    \label{eqn:Bsum-E}
    \\ 
    \Borel\LRbkL{ \frac{1}{\pi} \PIntZI{d\spr} \, 
    \frac{\Im \CorrE{l}{\Phen}{\spr,\va{p}} / \sqrt{\spr}}{\spr-s}} & = & 
    \Borel\LRbkL{\Re \CorrO{l}{\OPE}{s,\va{p}}},
    \label{eqn:Bsum-O}
\end{eqnarray}
where the Borel transform $\Borel$ is defined as 
\begin{\eqn}
    \Borel\LRbkL{\Pi(s)} = 
        \lim_{\tworow{-s,n\rightarrow\infty}{-s/n=\Msq}} 
        \frac{(-s)^{n}}{(n-1)!}\left (\frac{d}{ds}\right )^{n} \!\Pi(s),
\end{\eqn}
with $M$ being the Borel mass.

The left hand side of the sum rule (the phenomenological side) 
is assumed to have the pole + continuum structure: 
\begin{\eqn}
    \Corr{l\, ({\rm E, O})}{\Phen}{s,\va{p}} = 
        \Corr{l\, ({\rm E, O})}{\Phen\Pole}{s,\va{p}} 
        + \Corr{l\, ({\rm E, O})}{\Phen\Cont}{s,\va{p}},
\end{\eqn}
where the continuum part $\Corrp{l\, ({\rm E, O})}{\Phen\Cont}$ is 
extracted from $\Im\Corrp{l\, ({\rm E, O})}{\OPE}$, 
\begin{\eqn}
    \Im \Corr{l\, ({\rm E, O})}{\Phen\Cont}{s,\va{p}} = 
        \Im \Corr{l\, ({\rm E, O})}{\OPE}{s,\va{p}} 
            \theta \! \LRbk{s-\Thr^{\,l\, ({\rm E, O})}},
    \label{eqn:PhenCont}
\end{\eqn} 
with $\theta \! \LRbk{x}$ being the step function.
$\ThrE{\,l \,}$ and $\ThrO{\,l \,}$ are the continuum thresholds of 
$\Im \CorrEp{l}{\Phen}$ and $\Im \CorrOp{l}{\Phen}$ respectively.
The pole part for $\vect{p}=\vzero$ has been already discussed in 
Eqs.(\ref{eqn:ImCorrS-PhenM}) and (\ref{eqn:ImCorrV-PhenM}).
The right hand side of the sum rule (the OPE side) is 
derived from Eqs.(\ref{eqn:CorrSE-OPE}-\ref{eqn:CorrVO-OPE}).
For the reason which we denoted in the previous subsection, 
we use universal threshold $\Thr = \ThrE{\,l \,} = \ThrO{\,l \,}$.
However the height of the continuum is affected by 
the isospin-asymmetric nuclear medium, 
and $\Im \Corr{l}{\Phen\Cont}{\omg}$ becomes asymmetric 
under $\omg \leftrightarrow -\omg$.
This is taken into account in Eq.(\ref{eqn:PhenCont}) through 
the asymmetry of $\Im \Corr{l}{\OPE}{\omg}$. 
Its explicit form is shown 
in Eqs.(\ref{eqn:ImCorrS-OPEM}-\ref{eqn:ImCorrV-OPEM}).

For the scalar mixing angle, resulting BSRs from Eq.(\ref{eqn:Bsum-E}) 
at $\vect{p}=\vzero$ become 
\begin{\eqn}
    \BetaSL 
    \left( \MSz\,e^{-\MSzsq/\Msq} - \ML\,e^{-\MLsq/\Msq} \right) \MME{S} 
        = \ffunc{S}{\Msq,\Thr},
    \label{eqn:BSRM-SE}
\end{\eqn}
with
\[
    \ffunc{S}{\Msq,\Thr} \equiv 
    {\dispst 
        \CoeffSL \left[ \, \frac{1}{2} \left\{ 
        \dmm\, M^{6} \Efunc{2}{\frac{\Thr}{\Msq}}
        + (-4\pi^{2}) \EV{\diff{\qc{q}}} M^{4} \Efunc{1}{\frac{\Thr}{\Msq}} 
        \right\} \right]
    },
\]
and for vector mixing angle 
\begin{\eqn}
    \BetaSL 
    \left( \MSz\,e^{-\MSzsq/\Msq} - \ML\,e^{-\MLsq/\Msq} \right) \MME{V} 
        = \ffunc{V}{\Msq,\Thr},
    \label{eqn:BSRM-VE}
\end{\eqn}
with 
\[
    \ffunc{V}{\Msq,\Thr} \equiv 
    {\dispst 
        \CoeffSL \left[ 
        \frac{(4\pi^{2})}{4}\,\EV{\diff{\qn{q}}} M^{4} 
        \Efunc{1}{\frac{\Thr}{\Msq}} 
        \right]
    }.
\]
Here ${\Efunc{n}{x}=1-e^{-x}\sum^{n}_{r=0}x^{r}/r!}$.

One can simplify the above sum rules without loss of generality 
by expanding the left hand side in terms of 
a small parameter 
$\DM /M = (\MSz - \ML )/M \sim (\MSz - \ML )/((\MSz + \ML )/2) 
    \sim 77.0 \,{\MeV}/1.15 \,{\GeV} \sim 0.07$. 
Using this expansion and neglecting $O((\DM)^{2})$ contribution, 
Eq.(\ref{eqn:BSRM-SE}) becomes 
\begin{\eqn}
    \LRbk{\Msq - 2\Masq} \MME{S}_{{\Fn}} = \Ffunc{S}{\Msq,\Thr},
    \label{eqn:BSRS-F}
\end{\eqn}
where 
$\Ffunc{S}{\Msq,\Thr} \equiv 
    \ffunc{S}{\Msq,\Thr} \Msq e^{\Masq/\Msq} / (\BetaSL\,\DM )$ 
and we put the suffix {\Fn} to the mixing angle for later convenience.
One may alternatively take derivative of Eq.(\ref{eqn:BSRM-SE}) 
with respect to $\Msq$ to enhance the lower dimensional operator 
in OPE and then expand the result by $\DM/M$.
Then we obtain the second sum rule for the mixing angle 
\begin{\eqn}
    \LRbk{3\Msq - 2\Masq} \MME{S}_{{\Fd}} = \Ftfunc{S}{\Msq,\Thr},
    \label{eqn:BSRS-Fd}
\end{\eqn}
where 
$\Ftfunc{S}{\Msq,\Thr} \equiv ({d\!\,\ffunc{S}{\Msq,\Thr}}/{d\Msq}\,)
    ({\M^{6}}/{\Masq}\,)\,e^{\Masq/\Msq} / (\BetaSL\,\DM )$.

Extracting a term which is proportional to 
$\Msq$ and a term which is $\Msq$-independent in 
Eqs.(\ref{eqn:BSRS-F}) and (\ref{eqn:BSRS-Fd}), 
one finally arrive at four types of sum rules, which we call 
Type {\FP},{\FC},{\FdP} and {\FdC}. 
\begin{\eqn}
    \MME{S}_{\FP} = \frac{d}{d\Msq} \Ffunc{S}{\Msq,\Thr},
    \label{eqn:BSRS-FP}
\end{\eqn}
\begin{\eqn}
    \MME{S}_{\FC} = \left\{ \Msq \frac{d}{d\Msq} \Ffunc{S}{\Msq,\Thr} 
        - \Ffunc{S}{\Msq,\Thr} \right\}/(2\Masq),
    \label{eqn:BSRS-FC}
\end{\eqn}
\begin{\eqn}
    \MME{S}_{\FdP} = \frac{1}{3} \frac{d}{d\Msq} \Ftfunc{S}{\Msq,\Thr},
    \label{eqn:BSRS-FdP}
\end{\eqn}
and
\begin{\eqn}
    \MME{S}_{\FdC} = \left\{ \Msq \frac{d}{d\Msq} \Ftfunc{S}{\Msq,\Thr} 
        - \Ftfunc{S}{\Msq,\Thr} \right\}/(2\Masq).
    \label{eqn:BSRS-FdC}
\end{\eqn}

The Borel sum rules for the mixing angle $\MME{V}$ 
can be formulated exactly in the same manner starting from 
Eq.(\ref{eqn:BSRM-VE}).
The results are obtained by replacing the suffix $S$ by $V$ 
in Eqs.(\ref{eqn:BSRS-FP}), (\ref{eqn:BSRS-FC}), (\ref{eqn:BSRS-FdP}) 
and (\ref{eqn:BSRS-FdC}).

\setcounter{equation}{0}
\section{NUMERICAL RESULTS}
\label{sec:NumA}

In this section, we evaluate the absolute value of the mixing angle 
with the use of the BSRs, 
Type {\FP} (\ref{eqn:BSRS-FP}), {\FC} (\ref{eqn:BSRS-FC}), 
{\FdP} (\ref{eqn:BSRS-FdP}) and {\FdC} (\ref{eqn:BSRS-FdC}), 
supplemented with the FESRs (\ref{eqn:FESR-SE}) and (\ref{eqn:FESR-VE}).
To extract the mixing angles from the sum rules, 
we need to know various QCD parameters (vacuum condensates and quark masses), 
the coupling strength $\abs{\BetaSL}$ 
and also the Borel window in which Borel analysis is made.
They are determined by the following procedures.

\begin{enumerate}
    \item{QCD parameters and $\BetaSL$ in the vacuum} \\
        In Table \ref{tb:QCDparameter}, the QCD parameters 
        which we use in our analysis are summarized.
        These parameters reproduce 
        the mass spectrum of octet baryons in QCD sum rules 
        within 10\% \cite{NYM}.
        (OPE up to dimension 7 and quark masses up to the 2nd order 
        have been taken into account in this analysis.)
        $\BetaSL(=16\pi^{4}\csop{\Lambda}\csop{\Sigmaz})$ 
        has been determined by the BSR in the vacuum 
        for ``diagonal'' correlations ($A=B=\Lambda$ and $A=B=\Sigmaz$).  
        Using the parameters in Table \ref{tb:QCDparameter}, 
        we obtain $\abs{\BetaSL} = 2.5 \,{\GeV}^{6}$ 
        from the scalar-even sum rule \cite{NYM}.
        (Note that the sum rules for the diagonal correlations 
        provide only the absolute values of 
        $\csop{\Lambda}$ and $\csop{\Sigmaz}$.) 
        The optimum threshold $\Thr$ turns out to be $3.2 \,{\GeV}^{2}$ 
        from the Borel stability.
        This number is consistent with 
        the position of the second resonances of $\Lambda$ and $\Sigmaz$.
    
    \item{Mixing angle in the vacuum} \\
        The scalar-even BSR for $\szl$ mixing angle 
        in the vacuum $\AngleP_{\Vac}$
        (in which OPE up to dimension 7 and 
        quark mass up to 2nd order
        have been taken into account) 
        provides 
        $|\AngleP_{\Vac}|
        =1.4 \times 10^{-3} \LRbk{\frac{2.5\,{\GeV}^{6}}{|\BetaSL|}}$
        with the Borel window $1.4 \,{\GeV}^{2} \le \Msq \le 2.6 \,{\GeV}^{2}$
        and the threshold $\Thr \simeq 3.2 \,{\GeV}^{2}$ \cite{NYM}.
        Zhu {\it et al.} \cite{ZHY} obtained 
        $|\AngleP_{\Vac}|\simeq 7 \times 10^{-3} $ 
        for different QCD parameters 
        with $\abs{\BetaSL} = 1.76 \,{\GeV}^{6}$.
        We will discuss the effect of this difference to 
        the in-medium mixing angle at the end of this section.
    
    \item{QCD parameters and Borel window for in-medium mixing angle} \\
        The QCD parameter essential for obtaining 
        the in-medium mixing angles is Eq.(\ref{eqn:Mm052});
        \[
            {\rd(\dMN)}/{\rd(\dmm)} = 0.52.
        \]
        In the standard Borel analysis, 
        the Borel window is chosen such that 
        the higher orders in OPE and the continuum contribution 
        are well suppressed.
        The Borel window satisfying these conditions for 
        the $\szl$ mixing angle in the vacuum is 
        $1.4 \,{\GeV}^{2} \le \Msq \le 2.6 \,{\GeV}^{2}$ as shown above 
        where OPE up to dimension 7 
        has been taken into account \cite{NYM}.
        For the mixing angle in the medium, 
        we have OPE only up to dimension 4 
        in which the medium effects 
        appear only in the highest dimensional operators 
        and are dominant.
        Therefore, it is difficult 
        to find the Borel window and 
        to reach similar level of Borel stability.
        This is shown in Fig.\ref{fig:BSRmAngleS} 
        where in-medium mixing angle $\MME{S}_{\FdP}$ 
        as a function of the Borel mass 
        for different values of $\Thr$ is plotted.
        Since the Borel curve is not enough stable 
        in the medium, we simply adopt the Borel window 
        determined in the vacuum and extract the in-medium mixing angle 
        by making average over the Borel window. 
\end{enumerate}

\TbQCDparameter{htbp}
\FigBSRmAngleS{htbp}

Adopting the procedure described above, 
we obtain the in-medium mixing angles $\MME{S,V}$ 
as a function of the continuum threshold $\Thr$ 
for four different types of BSR; {Type \FP}, {\FC}, {\FdP} and {\FdC}.
They are shown in Fig.\ref{fig:BSRAngleS} and in Fig.\ref{fig:BSRAngleV} 
at nuclear saturation density 
$\DensT = 0.16\,{\rm fm}^{-3} \equiv \DensS$ and at 
typical value of the isospin-asymmetry for heavy nuclei such as Pb, 
$\dDens/\DensT=0.21$. 

\FigBSRAngleS{htbp}
\FigBSRAngleV{htbp}

 To reduce the uncertainties due to the absence of 
higher dimensional operators in OPE, 
we examine the reliability of each type of sum rules 
in the following ways.
First of all, if the BSRs are consistent with each other, 
$\MME{S,V}$ as a function of $\Thr$ should have a similar behavior 
between Type {\FP} and Type {\FC} and 
between Type {\FdP} and Type {\FdC}. 
Such comparison is shown 
in Fig.\ref{fig:BSRAngleS} and Fig.\ref{fig:BSRAngleV}.
The figures indicate that Type {\Fd} sum rules 
are more reliable than Type {\Fn} from this criterion.

One can make further selection of a reliable BSR 
by the comparison with FESR.
Remember that the $n$-th order term in the $1/\Msq$-expansion 
of the BSR is equivalent to the $n$-th order FESR.
Therefore, if OPE is well behaved, BSR and 
corresponding FESR should give the same result.
Such comparison is also shown 
in Fig.\ref{fig:BSRAngleS} and Fig.\ref{fig:BSRAngleV}.
 From the right panels of Fig.\ref{fig:BSRAngleS} and Fig.\ref{fig:BSRAngleV}, 
we conclude that the Type {\FdP} 
is more reliable than Type {\FdC} 
for reasonable range of the continuum threshold 
located around the second resonances of $\Lambda$ and $\Sigmaz$, 
$\Thr \simeq 3.2\,{\GeV}^{2}$.

In Fig.\ref{fig:BSRAngleFdP} we show 
the scalar angle $\MME{S}$ and the vector angle $\MME{V}$ 
in the Type {\FdP} as a function of the continuum threthold $\Thr$.
The curves in the Fig.\ref{fig:BSRAngleFdP} indicate 
the maximum (Max), the minimum (Min) and the average (Avg) value 
in the Borel window.
Fig.\ref{fig:BSRAsymmFdP} shows the $n-p$ asymmetry dependence of 
$\MME{S}$ and $\MME{V}$ 
for the total density $\DensT=0.5\,\DensS$, $\DensS$ and $1.5\,\DensS$.

\FigBSRAngleFdP{htbp}

Finally, by using the Type {\FdP} sum rule, 
we obtain the scalar and vector mixing angles as follows; 
\begin{eqnarray}
    \MME{S}_{\Med} & = & \LRbkL{-(0.19 \pm 0.02) \, \Asymm \, (\DensT/\DensS)} 
        \LRbk{\frac{2.5\,{\GeV}^{6}}{\BetaSL}} 
        \LRbk{\frac{3.9 {\MeV}}{\dmm}} ,
    \label{eqn:BSRAngleS(Medium)} 
    \\
    \MME{V}_{\Med} & = & \LRbkL{+(0.20 \pm 0.02) \, \Asymm \, (\DensT/\DensS)} 
        \LRbk{\frac{2.5\,{\GeV}^{6}}{\BetaSL}} ,
    \label{eqn:BSRAngleV(Medium)}
\end{eqnarray}
where the $n-p$ asymmetry is defined as $\Asymm \equiv \dDens/\DensT$.
The error bars are determined by 
the minimum and maximum values of 
the sum rule for mixing angles in the Borel window.
For typical values of the parameters, 
$\dmm = 3.9 \,{\MeV}$ and $\abs{\BetaSL} = 2.5 \,{\GeV}^{6}$, 
$\Dens{N} = \DensS$ and $\Asymm = 0.21$, 
$\MME{S}_{\Med}$ dominates over 
the vacuum mixing angle $\AngleP_{\Vac}$. 
Also, the relation $\MME{V}_{\Med}/\MME{S}_{\Med} \sim -1$ 
discussed in Sec.\ref{sec:QSR} is well satisfied.

\FigBSRAsymmFdP{htbp}

The particle and anti-particle mixing angles 
$\AngleP$ and $\AngleA$ are obtained from 
Eq.(\ref{eqn:AngleP}) and (\ref{eqn:AngleA}) as 
\begin{eqnarray}
    \!\!\!\!\!\!\!\! 
    \AngleP & = & 
        [(0.01 \pm 0.04) \, \Asymm \, (\DensT/\DensS)] \, \sgn{\BetaSL} 
        + \AngleP_{\Vac}, 
    \label{eqn:BSRAngleP(Medium)} 
    \\ 
    \!\!\!\!\!\!\!\! 
    \AngleA & = & 
        [(-0.39 \pm 0.04) \, \Asymm \, (\DensT/\DensS) ] \, \sgn{\BetaSL}
        + \AngleP_{\Vac},
    \label{eqn:BSRAngleA(Medium)}
\end{eqnarray}
with $\dmm = 3.9 \,{\MeV}$, $\abs{\BetaSL} = 2.5 \,{\GeV}^{6}$. 

In the analysis of the $\szl$ mixing in the vacuum 
by Zhu {\it et al.} \cite{ZHY}, 
they use different set of QCD parameters from Table \ref{tb:QCDparameter} 
(in particular $\dmm=3.0 \,{\MeV}$ and 
$\EVV{\ave{\qc{q}}}=(-241\,{\MeV})^{3}$) 
and obtain $\abs{\BetaSL}=1.76 \,{\GeV}^{6}$ with 
the threshold $\Thr=3.4 \,\GeV^{2}$ and 
the Borel window $1.3 \,{\GeV}^{2}$ $\le$ $\Msq$ $\le$ $2.5 \,{\GeV}^{2}$.
Substituting these values into 
Eqs.(\ref{eqn:BSRAngleS(Medium)}) and (\ref{eqn:BSRAngleV(Medium)}), 
one obtains 
\begin{eqnarray}
    \AngleP & = & 
        [(-0.12 \pm 0.08) \, \Asymm \, (\DensT/\DensS)] \, \sgn{\BetaSL} 
        + \AngleP_{\Vac}, 
    \label{eqn:BSRAngleP(Medium)Z}
    \\ 
    \AngleA & = & 
        [(-0.75 \pm 0.09) \, \Asymm \, (\DensT/\DensS)] \, \sgn{\BetaSL}
        + \AngleP_{\Vac}.
    \label{eqn:BSRAngleA(Medium)Z}
\end{eqnarray}
which are qualitatively consistent with the result 
obtained using our parameter set.

\section{CONCLUSIONS}
\setcounter{equation}{0}
\label{sec:Conclud}

In this paper, we have studied the $\szl$ mixing angles 
in the isospin-asymmetric nuclear medium by using the QCD sum rules.

Firstly, we have discussed general properties of 
diagonal and off-diagonal correlation functions of the baryonic currents.
We found that the off-diagonal (mixed) correlation function consists of 
scalar, vector and tensor terms.
They are further decomposed into even and odd parts 
in terms of the reflection symmetry under 
$\omega \leftrightarrow - \omega$.
Then we derived dispersion relations for each component.

Secondly, we examined the general structure of the mixing angle for baryons 
and introduced two independent mixing parameters $\MME{S}$ and $\MME{V}$ 
for the baryon at rest inside the medium.
The sum (difference) of these parameters are shown to be 
the particle mixing angle $\AngleP$ 
(the anti-particles mixing angle $\AngleA$).
This situation is analogous to 
the self-energy of the nucleon and anti-nucleon 
in the relativistic mean-field theories.

Thirdly, we have carried out the OPE for 
the $\szl$ mixed correlation function.
Then we constructed sum rules for $\MME{S}$ and $\MME{V}$.
 From the finite energy sum rules, we found that 
$\MME{V}_{\Med}/\MME{S}_{\Med} \sim -1$.
This implies that 
the particle mixing angle $\AngleP \,(= \MME{S} + \MME{V})$ in the medium 
is nearly equal to the one in the vacuum, and 
the isospin-asymmetric medium affects mainly 
the anti-particle mixing $\AngleA \,(= \MME{S} - \MME{V})$.
 From the Borel sum rules, we evaluated the in-medium parts of 
$\AngleP$ and $\AngleA$ numerically.
The results are summarized in 
Eqs.(\ref{eqn:BSRAngleP(Medium)}-\ref{eqn:BSRAngleA(Medium)Z}) 
in Sec.\ref{sec:NumA}.
As the baryon density and the isospin-asymmetry of the medium increase, 
the anti-particle mixing is enhanced, 
while the particle mixing remains less than 20\% 
of the anti-particle mixing.

The strong correlation between 
$\MME{S}_{\Med}$ and $\MME{V}_{\Med}$
and the strong modification of the 
anti-particle mixing in the isospin-asymmetric medium 
shown in this paper are 
model independent consequence supported 
both by the finite energy sum rules and the Borel sum rules.
On the other hand, the absolute magnitude of each mixing angle 
has uncertainties due to the absence of higher dimensional operators in OPE.
Better evaluation of the matrix elements of 
isospin-asymmetric operators beyond dimension 4 
is necessary for precise determination of the mixing angles.
Also, it is an open but interesting problem to study 
whether one can measure 
the anti-particle mixing in nuclei in the laboratory experiments.

\section*{ACKNOWLEDGEMENTS}
\vspace{-0.1mm}
N.Y. thanks to Prof. T.Kohmura for his encouragement
and to the members of Nuclear Theory Group 
at Kyoto University and University of Tokyo 
for their hospitality during the course of this work.

\appendix
\section*{APPENDIX}
\setcounter{section}{1}
\setcounter{equation}{0}
\label{App:Dim4Cond}

In this Appendix, we evaluate a dim.4 isospin anti-symmetric condensate 
$\EV{\diff{\qcv{q}{\,i\zero{\cd}\zero{\gm}}}}$ following \cite{DHP}.

In the medium, the dim.4 quark condensate 
$\EV{\qcv{q}{\,i\zero{\cd}\zero{\gm}}}$ is represented as 
\begin{\eqn}
    \EV{\qcv{q}{\,i\zero{\cd}\zero{\gm}}} = 
        \frac{1}{4}\, m_{q} \EV{\qc{q}} 
        + \EV{\qcv{q}{\,i\!\LRbk{\zero{\cd}\zero{\gm} - \slacd/4}}}.
\end{\eqn}
The vacuum part of the second term vanishes 
\begin{\eqn}
    \EV{\qcv{q}{\,i\!\LRbk{\zero{\cd}\zero{\gm} - \slacd/4}}} 
    \simeq \EVs{\qcv{q}{\,i\!\LRbk{\zero{\cd}\zero{\gm} - \slacd/4}}}{p} 
        \Dens{p} 
    + \EVs{\qcv{q}{\,i\!\LRbk{\zero{\cd}\zero{\gm} - \slacd/4}}}{n} 
        \Dens{n}. 
\end{\eqn}
The expectation value taken by the proton at rest is \cite{HL} 
\begin{\eqn}
    \EVs{ \qcv{q}{\,i\!\LRbk{\zero{\cd}\zero{\gm} - \slacd/4}} }{p} = 
        \frac{3}{8}\, \M_{p} A^{q}_{2}(\mu^{2})
\end{\eqn}
where $A^{q}_{2}(\mu^{2})$ is the 2nd moment of 
the parton distribution function $q\,(x,\mu^{2})$, $\qb\,(x,\mu^{2})$
of the proton 
\begin{\eqn}
    A^{q}_{n}(\mu^{2}) = 2 \int^{1}_{0} dx\, x^{n-1} 
        \LRbkM{q\,(x,\mu^{2})+(-1)^{n-1}\qb\,(x,\mu^{2})}.
    \label{eqn:PDFP}
\end{\eqn}
For isospin anti-symmetric condensate 
$\EV{\diff{\qcv{q}{\,i\zero{\cd}\zero{\gm}}}}$, 
we obtain 
\begin{\eqn}
    \EVs{ \diff{\qcv{q}{\,i\!\LRbk{\zero{\cd}\zero{\gm} - \slacd/4}}} }{p} 
        = \frac{3}{8}\, \M_{p} \diff{A^{q}_{2}(\mu^{2})}.
\end{\eqn}
Then, the expectation value taken by the neutron at rest becomes 
\begin{\eqn}
    \EVs{ \diff{\qcv{q}{\,i\!\LRbk{\zero{\cd}\zero{\gm} - \slacd/4}}} }{n} 
        = - \frac{3}{8}\, \M_{n} \diff{A^{q}_{2}(\mu^{2})} 
\end{\eqn}
up to the 1st order in isospin-asymmetry.
Thus we finally arrive at 
\begin{eqnarray}
    \!\!\!\!\!\!\!\!
    \EV{\diff{\qcv{q}{\,i\zero{\cd}\zero{\gm}}}} \simeq 
        \frac{1}{4} \LRbkM{\ma \EV{\diff{\qc{q}}} 
        + \dmm \EV{\widehat{\qc{q}}}} 
        - \frac{3}{8}\, \diff{A^{q}_{2}(\mu^{2})} 
        \,(\MaN \, \dDens + \dMN \, \DensT/2).
\end{eqnarray}


\end{document}